\documentclass[12pt,a4paper]{article}\usepackage{latexsym,amssymb,amsmath,epsfig,graphicx}

\newcommand{\pkt}{\; .} 
\newcommand{\kma}{\; ,}

\newcommand{\beast}{\begin{eqnarray*}}
\newcommand{\eeast}{\end{eqnarray*}}
\newcommand{\eqn}[1]{(\ref{#1})}

\newcommand{\calf}{{\cal F}}

\newcommand{\call}{{\cal L}}

\newcommand{\calm}{{\cal M}}

\newcommand{\calv}{{\cal V}}

\newcommand{\bfx}{{\bf x}}

\newcommand{\bs}{\boldsymbol}
\newcommand{\be}{\begin{equation}}
\newcommand{\ee}{\end{equation}}
\newcommand{\bea}{\begin{eqnarray}}
\newcommand{\eea}{\end{eqnarray}}

\begin{document}
\noindent
DESY 10-079   \\
June 2010

\vspace{8mm}
\begin{center}
{\Large \bf Coupled scalar fields in a flat FRW
 universe: renormalisation}
\\\vspace*{8mm}
{\large J\"urgen Baacke$^{\rm 1}$
\let\thefootnote\relax\footnotetext{E-mails:~~juergen.baacke@tu-dortmund.de},
Laura Covi$^{\rm 2}$\footnote{\hspace*{1.5cm}laura.covi@desy.de}, 
Nina Kevlishvili$^{\rm 2,3}$\footnote{\hspace*{1.5cm}nina.kevlishvili@desy.de}
\\[3mm]
\normalsize
$^{\rm 1}$ Fakult\"at Physik, Technische Universit\"at Dortmund \\
D - 44221 Dortmund, Germany\\
\normalsize
$^{\rm 2}$ Deutsches Elektronen Synchrotron DESY \\ D - 22603 Hamburg,
Germany, \\
\normalsize
$^{\rm 3}$ Andronikashvili Institute of Physics, GAS, 0177 Tbilisi, Georgia
}\\
\vspace{5mm}
\end{center}

\begin{abstract}
We study the non-equilibrium dynamics of a system of coupled scalar
fields in a Friedmann-Robertson-Walker (FRW) universe. 
We consider the evolution of spatially homogeneous ``classical'' 
fields and of their quantum fluctuations including the quantum
backreaction in the one-loop approximation. 
We discuss in particular the dimensional regularisation of the 
coupled system and a special subtraction procedure in order 
to obtain the renormalised equations of motion and the 
renormalised energy-momentum tensor and ensure that the energy
is well-defined and covariantly conserved. 
These results represent at the same time a theoretical
analysis and a viable scheme for stable numerical simulations. 
As an example for an application of the general formalism,
we present simulations for a hybrid inflationary model. 
\end{abstract}


\section{Introduction}

\label{sec:introduction}

While the renormalisation of a single scalar field in a 
Friedmann-Robertson-Walker (FRW) universe, and fermion fields 
coupled to such field, have been discussed by various 
authors~\cite{Ringwald:1986wf,Boyanovsky:1993xf,Boyanovsky:1997xt,
Boyanovsky:1997cr,Ramsey:1997sa,Baacke:1999nq}, 
a consistent and coherent formulation of coupled scalar field models 
including non-minimal gravity couplings and full renormalisation 
is still missing. As a first step in that direction, we study
here such a system of coupled fields in the one-loop approximation
with special emphasis on its renormalisation in the $ \overline{MS} $
scheme. This will allow us to obtain a consistent set of coupled
equations of motion without divergences and numerical 
instabilities, which is highly suitable for numerical applications.
Coupled scalar fields appear in cosmology in multi-field models 
of inflation and are important as well in particle physics in 
the context of grand unified theories, which contain several 
Higgs fields. 

The standard example of a coupled system of scalar fields
in cosmology is of course the hybrid model of inflation
which has been introduced by Linde 
\cite{Linde:1990gz,Linde:1991km,Linde:1993cn} and whose coupled
dynamics at the end of the inflationary phase has received wide 
attention~\cite{Garcia-Bellido:1997wm,Lyth:1998xn,Garcia-Bellido:1999sv,
Bastero-Gil:1999fz,Krauss:1999ng,Felder:2000hj,Micha:1999wv,Felder:2001kt,
Copeland:2001qw,Buchmuller:2000zm,Nilles:2001fg,Asaka:2001ez,
Cormier:2001iw,Garcia-Bellido:2002aj,Borsanyi:2002tm,Borsanyi:2003ib}
(for a recent review see \cite{Allahverdi:2010xz}).
Unlike other models for inflation and for preheating, where one
has just one ''inflaton'' field with vacuum quantum numbers
 coupled to various other boson and fermion fields, this type
of models has two scalar fields with can acquire a time dependent 
vacuum expectation value. So these fields and their quantum
fluctuations can mix dynamically. 

The renormalisation of such a coupled system of bosons out of equilibrium
has been considered previously by Cormier et al. \cite{Cormier:2001iw} 
in the one-loop approximation, and in Ref. \cite{Baacke:2003bt}
in the Hartree approximation. The one-loop approximation includes 
the quantum back-reaction of the quantum fluctuations on the 
classical fields, the Hartree approximation includes in addition
the backreaction on the quantum fluctuation themselves.
This approximation has been considered as well by Bastero-Gil
et al. \cite{Bastero-Gil:1999fz} with a momentum cutoff for the 
quantum fluctuations.   
These publications do not include the coupling of the fields to 
gravity, which may be a reasonable approximation in the period of 
preheating after inflation. However, when describing the evolution 
during inflation and the end of inflation, the inclusion of the 
coupling to gravity could be important and it is necessary for
consistency, since those couplings are generated by quantum 
corrections. 
Moreover in recent years, such non-minimal gravitational coupling
has drawn a lot of attention in the context of realising inflation
within the Standard Model \cite{Bezrukov:2007ep,Kaloper:2008gs}
and it is surely an interesting issue to include it also in the 
case of many fields. In those models it became clear pretty soon
that quantum effects on the inflation potential cannot be neglected
\cite{DeSimone:2008ei,Bezrukov:2008ej, GarciaBellido:2008ab}
and it may be the same also for the hybrid case.
 
The one-loop renormalisation for the out-of-equilibrium evolution 
of a single scalar field in a FRW universe has been considered 
by several authors~\cite{Ringwald:1986wf,Boyanovsky:1993xf,
Ramsey:1997sa,Bordag:1998zj,Lindig:1998qh,
Baacke:1997rs,Baacke:1999gc}. The technical approach which we will 
use here is based on Ref.~\cite{Baacke:1999gc}, which itself uses
a formalism developed in Ref.~\cite{Baacke:1996se}.
As in these references we establish here a ''renormalised computation 
scheme'', i.e., the renormalisation is taken into account with the 
same rigour as in more formal approaches, and at the same time the 
formalism can be implemented efficiently into a numerical code.
In particular the renormalisation is independent of the initial 
conditions, though the divergent momentum integrals depend on the 
initial effective masses of the system. The energy-momentum tensor 
is in this formulation always finite and covariantly conserved.

The paper is organised as follows: in Sec. \ref{sec:shortreview}
we recall some basic relations of FRW cosmology, with special 
attention to an extension beyond space-time dimension $n=4$ and
with the inclusion of higher dimensional gravity counterterms, 
which are needed in order to properly renormalise the coupled
fields system; in Sec. \ref{sec:themodel} we define the general 
model whose dynamical equations we will formulate in the one-loop 
approximation; in Sec. \ref{sec:renormalization} we discuss
in detail the regularisation and renormalisation of the
equation of motions and of the energy-momentum tensor.
Finally in Sec.~\ref{sec:simulations} we give few selected
results to show the applicability of this scheme and the
stability of the numerical results and then conclude. More 
exhaustive numerical results will be given in a forthcoming
publication. Various technical details can be found in the 
Appendices~\ref{app:ndimtensors}-\ref{app:anexample}.


\section{Short review of FRW  cosmology}
\setcounter{equation}{0}\label{sec:shortreview}

We consider here a spatially isotropic and flat space-time
with $n-1$ spacial dimensions, which can be described by a FRW 
metric with $k=0$ curvature parameter as
\begin{equation}
ds^2=dt^2-a^2(t)d\bs{x}^2\; ,
\end{equation}
where $ d\bs{x}^2 = \sum_{i=0}^{n-1} dx_i^2 $.
The effective action of the coupled gravity-scalar fields system
is given by
\begin{eqnarray}\label{eq:action}
{\cal S} &=& 
\int d^{n} x \sqrt{-g} \left[\frac{M_P^2}{2} R + \Lambda g_{\mu\nu}
+ \delta\alpha R^2 +\delta\beta R\,^{\alpha\beta}R\,_{\alpha\beta}
+\delta\gamma R\,^{\alpha\beta\gamma\delta} R\,_{\alpha\beta\gamma\delta}
\right] \nonumber\\
& & + \int d^n x\; {\cal L} (\Phi_i,g_{\mu\nu}, R) \pkt
\end{eqnarray}
The first two terms are the Einstein-Hilbert action and a cosmological 
constant term, which allows to tune the energy density of the ground 
state to zero. We have introduced also all the gravitational terms up 
to dimension four for later convenience.
The last term is instead given by the action for the matter 
fields, in our case a system of coupled bosons described in the
next section.

The time evolution of the $a(t)$ is governed by the
Einstein's field equation, which for the Einstein-Hilbert 
case $ \delta\alpha = \delta \beta = \delta \gamma = 0 $
is given by
\begin{equation} \label{eq:Einstein}
G_{\mu\nu} +\Lambda g_{\mu\nu}=-8\pi G<T_{\mu\nu}>\pkt
\end{equation}
where the general expression for energy-momentum tensor is 
\begin{equation}\label{eq:energmomentumdef}
T_{\mu\nu}(x)=\frac{2}{\sqrt{-g(x)}}
\frac{\delta {\cal S}_{matter}}{\delta g_{\mu\nu}}\pkt
\end{equation}

In quantum field theory renormalisation requires to introduce
a counterterm action. Besides the usual mass and coupling constant 
counterterms the coupling to gravity induces divergences which 
require the inclusions of the higher curvature terms written in
Eq.~(\ref{eq:action}) and therefore a modification of the original 
Einstein field equations.
These then take the form
\begin{equation} \label{eq:Einsteinrenorm}
(1+\delta Z)G_{\mu\nu}+\delta\alpha ^{(1)}H_{\mu\nu}
+\delta\beta ^{(2)}H_{\mu\nu}+\delta\gamma H_{\mu\nu}
+(\Lambda + \delta\Lambda) g_{\mu\nu}=-8\pi G<T_{\mu\nu}>\pkt
\end{equation}
The  tensors  $^{(1)}H_{\mu\nu}, ^{(2)}H_{\mu\nu}$ and
$ H_{\mu\nu}$ are related to the variation of 
the higher curvature 
terms $R^2,~ R^{\alpha\beta}R_{\alpha\beta}$ and 
$R^{\alpha\beta\gamma\delta}R_{\alpha\beta\gamma\delta}$. 
Since gravity is not quantised, all divergences arise, at least technically, 
from the quantum fluctuations of the matter fields in the 
energy-momentum tensor. Therefore one may prefer to include the 
counterterms into a redefinition of the right hand side of the 
Einstein equations via
\be \label{eq:energymomentumtensorredef}
 T_{\mu\nu} \rightarrow T_{\mu\nu}+\delta \tilde Z
G_{\mu\nu}+\delta\tilde \alpha ^{(1)}H_{\mu\nu}
+\delta \tilde \beta ^{(2)}H_{\mu\nu}+\delta\tilde
\gamma H_{\mu\nu} +(\tilde \Lambda + \delta\tilde \Lambda) g_{\mu\nu}
\pkt\ee
Here $\delta \tilde Z= \delta Z/8\pi G$ and similarly for the
other terms.

For the various tensors we use the 
conventions of Ref. \cite{Birrell:1982ix}:
the Einstein curvature tensor is defined as
\begin{equation}
G_{\mu\nu}=R_{\mu\nu}-\frac{1}{2}g_{\mu\nu}R\,;
\end{equation}
the Ricci tensor and the Ricci scalar are 
\begin{equation}
R_{\mu\nu}=R^\lambda_{\mu\nu\lambda},\hspace*{1cm} R=g^{\mu\nu}R_{\mu\nu}
\kma\end{equation}
where the curvature tensor is given by
\begin{equation}
R_{\alpha\beta\gamma}^\lambda=\partial_\gamma\Gamma_{\alpha\beta}^\lambda-\partial_\alpha\Gamma_{\gamma\beta}^\lambda
+\Gamma_{\gamma\delta}^\alpha\Gamma_{\alpha\beta}^\delta-\Gamma_{\alpha\delta}^\lambda\Gamma_{\gamma\beta}^\delta \pkt
\end{equation}

The higher curvature tensors are defined as
\cite{Birrell:1982ix}
\bea
^{(1)}H\,_{\mu\nu} &&=\frac{1}{\sqrt{-g}}\frac{\delta}{\delta g\,^{\mu\nu}}
\int\!d^nx \sqrt{-g}\,R^2\nonumber\\
&&=2\,R_{;\mu\nu}-2g\,_{\mu\nu}\;\Box R
-\frac 1 2 g\,_{\mu\nu}\,R^2+2RR\,_{\mu\nu}\kma\\
^{(2)}H\,_{\mu\nu}&&=\frac{1}{\sqrt{-g}}\frac{\delta}{\delta g\,^{\mu\nu}}
\int\!{d}^nx \sqrt{-g}\,R\,^{\alpha\beta}R\,_{\alpha\beta}\nonumber\\
&&=2 R\,^{\alpha}_{\mu\, ;\nu\alpha}- 
\Box R_{\mu\nu}-\frac 1 2g\,_{\mu\nu}\,\Box R
+2R\,_{\mu}^{\alpha}R\,_{\alpha\nu}-
\frac{1}{2}g\,_{\mu\nu}\,R\,^{\alpha\beta}
R\,_{\alpha\beta}\nonumber\\
&&=R\,_{;\mu\nu}-\frac{1}{2}g\,_{\mu\nu}\,\Box R-
\Box R\,_{\mu\nu}-\frac 1 2 g\,_{\mu\nu}
\,R\,^{\alpha\beta}R\,_{\alpha\beta}
+2R\,^{\alpha\beta}R\,_{\alpha\mu\beta\nu}\kma \nonumber\\\\
H\,_{\mu\nu}&&=\frac{1}{\sqrt{-g}}\frac{\delta}{\delta g\,^{\mu\nu}}
\int\!{d}^nx \sqrt{-g}\,R\,^{\alpha\beta\gamma\delta}
R\,_{\alpha\beta\gamma\delta}\nonumber\\
&&=
-\frac 1 2 g\,_{\mu\nu}R\,^{\alpha\beta\gamma\delta}
R_{\alpha\beta\gamma\delta}+2
R\,^{\mu\alpha\beta\nu}R\,_{\nu}^{\alpha\beta\gamma}
-4\Box R\,_{\mu\nu}+2R\,_{;\mu\nu}\nonumber\\
&&
-4R\,_{\mu\alpha}R\,^{\alpha}_{\nu}+4R\,^{\alpha\beta}
R\,_{\alpha\mu\beta\nu}\pkt
\eea
For the case $n=4$ in conformally flat space-time one has
\begin{equation} \label{eq:relationforHsin4D}
H_{\mu\nu}={}^{(2)}H_{\mu\nu}=\frac{1}{3} {}^{(1)}H_{\mu\nu}
\pkt\end{equation}

The explicit expressions for all these tensors in FRW space-time
are of course well-known. However, we will use dimensional 
regularisation, which implies that we have to perform all 
the computations for general $n=4-\epsilon$.
We therefore recall the explicit formulae for the relevant
tensors in flat $n$ dimensional FRW geometry in Appendix
\ref{app:ndimtensors}.

For the spatially isotropic FRW universe the Einstein field 
equations reduce for the time-time component 
and trace to the Friedmann equations
\begin{eqnarray}
G_{tt}&=&-8\pi GT_{tt}=-8\pi G \rho\kma\\
G_\mu^\mu&=&-8\pi GT_\mu^\mu=-8\pi G (\rho - (n-1) p)\kma
\end{eqnarray}
where $\rho$ is the energy density, and the pressure $p$
is given by 
\be
p=(\rho-T_\mu^\mu)/(n-1)\pkt
\ee.
The covariant conservation of energy takes then the form 
\be \label{eq:covencons}
\frac{d\rho}{dt}=(n-1)H(\rho+p)=H(n \rho -T_\mu^\mu)
\pkt\ee

\section{Coupled fields in curved space-time}
\setcounter{equation}{0}\label{sec:themodel}

We here consider the quantum field theory of $N$ coupled scalar 
fields. More specifically we consider a class of models whose 
Lagrangian is given by
\begin{equation}\label{curvedlagrangian}
\call=\sqrt{-g}\left\{
\frac{1}{2}\sum_i g^{\mu\nu}\partial_\mu\Phi_i\partial_\nu\Phi_i-
V(\Phi)-W(R,\Phi)\right\}\kma
\end{equation}
with the potential 
\begin{equation}\label{eq:potential}
V(\Phi)=\frac{1}{2}\sum_{i=1}^N m_i^2\Phi_i^2+
\frac{1}{4}\sum_{i=1}^N\sum_{j=1}^N\lambda_{ij}\Phi_i^2
\Phi_j^2 \pkt
\end{equation}
Instead $W(R,\phi)$ describes the coupling of the scalar fields 
to the curvature scalar:
\begin{equation}
W(R,\Phi)=\sum_i\frac{\xi_i}{2}R\;\Phi_i^2\pkt
\end{equation}
The  $\xi_i$  are dimensionless coupling constants that have no counterpart
 in the flat space theory. As we will see later there is an obviously
preferred choice $\xi_i=1/6$, called conformal coupling. 
The value $\xi_i=0$ is called instead minimal coupling.

\subsection{The equation of motion in the one-loop approximation}

We split the fields in two parts, the expectation values and the 
quantum fluctuations around it
\be
\Phi_i=\phi_i(t)+\psi_i(t,\bs{x})\kma
\ee
where the {\em classical field} $\phi_i(t)$ is assumed to be homogeneous
in space. 
The classical part of the Lagrangian density retains the form
\begin{eqnarray}\nonumber
\call^{(0)}&=&\sqrt{-g}\left\{
\frac{1}{2}\sum_i \left[ g^{\mu\nu}\partial_\mu\phi_i\partial_\nu\phi_i
-(m_i^2+\xi_i R) \phi_i^2\right]\right.\\
&&\left. \hspace{20mm}-\frac{1}{4}\sum_{ij}\lambda_{ij} \phi_i^2
\phi_j^2-\tilde\Lambda\right\} \pkt
\end{eqnarray}

The first order in the fluctuations vanishes (the expectation value of the fluctuations is zero). The fluctuation Lagrangian density of the second order in fluctuations
\bea\nonumber
\call^{(2)}&=&\sqrt{-g}\left\{\sum_i\frac{1}{2}g^{\mu\nu}\partial_\mu
\psi_i\partial_\nu\psi_i
-\frac{1}{2}\sum_i (m_i^2+\xi_i R)\psi_i^2\right.\\
&&\left.\hspace{15mm}-\frac{1}{2}\sum_{ij}\lambda_{ij}(\phi_i^2\psi_j^2+
2\phi_i\phi_j\psi_i\psi_j)\right\}\pkt
\eea
Now we can write the equations of motion for classical fields:
\bea\nonumber
&&\ddot\phi_i+(n-1)H\dot\phi_i+(m_i^2+\xi_i R)\phi_i
 \\
 &&\hspace{15mm}+\sum_j\lambda_{ij}\left[(\phi_j^2+<\psi_j^2>)\phi_i+
2<\psi_i\psi_j>\phi_j\right]=0\kma
\eea
and for the quantum fluctuations in the one-loop approximation:
\be
\ddot\psi_i+(n-1)H\dot\psi_i
+\frac{1}{a^2}\nabla^2\psi_i+(m_i^2+\xi_i R)\psi_i
+ \sum_j\lambda_{ij}\left( 2\phi_i\phi_j\psi_j+
\phi_j^2 \psi_i\right)
=0\pkt
\ee
The latter equation can be written in the form
\be
\ddot\psi_i+(n-1)H\dot\psi_i
+\frac{1}{a^2}\nabla^2\psi_i+\sum_j\calm^2_{ij}\psi_j=0
\kma\ee
where the mass squared matrix is given by
\be
\calm_{ij}^2=(m_i^2+\xi_i R)\delta_{ij}+2\lambda_{ij}\phi_i\phi_j
+\delta_{ij}\sum_k\lambda_{ik}\phi_k^2\kma
\ee
or, more explicitly,
\bea
\calm_{ii}^2&=&m_i^2+\xi_i R+3 \lambda_{ii}\phi_i^2 
+\sum_{k\neq i}\lambda_{ik}\phi_k^2  \kma
\\
\calm_{ij}^2&=&2\lambda_{ij}\phi_i\phi_j\hspace{10mm}i\neq j
\pkt
\eea
The expectation values of the 
quantum fluctuations can be expressed in terms of  
equal-time Green's functions via
\be
<\psi_i\psi_j>=-iG_{ij}(t,x;t,x)\kma
\ee
where the Green's functions satisfy 
\bea\nonumber
&&\hspace{10mm}\left\{\left[\frac{\partial^2}{\partial t^2}
+(n-1)H\frac{\partial}{\partial t}
+\frac{1}{a^2}\nabla^2\right] \delta_{ij}
+\calm_{ij}^2(t)\right\}G_{jk}(t,\bs{x};t',\bs{x}')
\\&&\hspace{20mm}=
\delta_{ik}\;\frac{i}{a^{n-1}}\;\delta^{n-1}(\bs{x}-\bs{x}')\;\delta(t-t')\pkt
\label{eq:Greeneqsnonherm}\eea

We have introduced here the set of equations for the classical fields 
and for the fluctuations in a heuristic way. A rigorous derivation
can be found in Ref. \cite{Calzetta:1986ey}, based on the Schwinger-Keldysh 
or closed-time path formalism \cite{Schwinger:1960qe,Keldysh:1964ud}.

The differential operator in Eq. \eqn{eq:Greeneqsnonherm} is
non-Hermitian due to the term $(n-1)H\partial/\partial t$, and so 
is the matrix valued  Green' s function $G$. This 
problem can be solved by introducing conformal time
\begin{equation}
\tau=\int_0^tdt'\frac{1}{a(t')} \pkt 
\end{equation}
We can then introduce the dimensionless rescaled fields and their fluctuations 
\begin{eqnarray}
\phi_i(t)&=&a^{-n/2+1}\tilde\phi_i(\tau)\kma\nonumber\\
\psi_i(t,\bfx)&=&a^{-n/2+1}\tilde\psi_i(\tau,\bfx)\pkt\nonumber
\end{eqnarray}
as well as the rescaled Green's function
\begin{equation}
\tilde{G}_{ij}(\tau,\bfx;\tau',\bfx')
=a^{n/2-1}(t)\;a^{n/2-1}(t')\;G_{ij}(t,\bfx;t',\bfx')
\pkt\end{equation}
As a function of conformal time, the Hubble parameter, the Ricci scalar 
and the derivatives of the fields are given by
\begin{eqnarray}
H&=&\frac{a'}{a^2}\kma\nonumber\\
R&=&2(n-1)\frac{a''}{a^3}+(n-1)(n-4)H^2\kma\nonumber\\
\dot{f}&=&a^{-n/2}\left[\tilde{f}'-\frac{n-2}{2}a H \tilde{f}\right]\kma
\\\nonumber
\ddot{f}&=&a^{-n/2-1}\left[{\tilde f\,}''-(n-1)aH\tilde{f}'
+\frac{(n-1)(n-2)}{2}a^2H^2\tilde f\right.\\
\nonumber &&\hspace{30mm}\left.
-a^2\frac{n-2}{4(n-1)}R\tilde f\right] \kma
\end{eqnarray}
where $f$ stands for any field and the prime denotes 
the derivative with respect to conformal time.

Then the equations of motion of the classical fields become
\be
\tilde{\phi}_i''+
\left(m_i^2+(\xi_i-\xi_n)R\right)a^2 \tilde \phi_i
+a^{4-n}\sum_j\lambda_{ij}\left[(\phi_j^2+\tilde G_{jj})\phi_i+
2\tilde G_{ij}\phi_j\right]=0\kma
\ee
where
\be
\xi_n = \frac{n-2}{4(n-1)}
\pkt\ee
We see here that for $ n=4 $ and conformal coupling $ \xi_i = 1/6 $,
the curvature term disappears from the equation of motion of the
fields, which reduces to the form it has in flat Minkowski space
if also $ m_i = 0$. 

The new two-point functions satisfy now
\begin{equation}
\left[\left(\frac{\partial^2}{\partial\tau^2}-\nabla^2\right)\delta_{ij}
+\tilde{\calm}^2_{ij}(\tau)\right]\tilde{G}_{ij}=
-\delta(\tau-\tau')\delta^{n-1}(\bfx-\bfx')\kma
\end{equation}
with the effective masses of the rescaled fluctuation fields
\begin{eqnarray}
\tilde{\calm}^2_{ii} (\tau) &=&
\left(m_i^2+(\xi_i-\xi_n)R\right)a^2
+a^{4-n}\left(3\lambda_{ii}\tilde{\phi_i}^2
+\sum_{j\neq i}\lambda_{ij}\tilde\phi_j^2\right)\kma\\
\tilde{\calm}^2_{ij} (\tau) &=&2a^{4-n}\lambda_{ij}\tilde{\phi_i}
\tilde{\phi_j}\hspace{10mm}i \neq j\pkt
\end{eqnarray}
The equations for the fluctuation fields in conformal time become
\be
\tilde{\psi_i}''+\nabla^2\tilde{\psi_i}+\sum_j
\tilde{\calm}^2_{ij} (\tau) \tilde{\psi_j}=0\pkt
\ee
We expand the fluctuation fields in terms of mode functions $f_i^\alpha(\tau;\bs{k})$
\be
\tilde\psi_i(\tau,\bs{x})=
\int\frac{d^{n-1} k}{(2\pi)^{n-1}}e^{i\bs{kx}}f_i^\alpha(\tau;\bs{k})\kma
\ee
which satisfy the mode equations
\begin{equation}\label{eq:modeeq}
{f_i^{\alpha}}''(\tau;\bs{k})+k^2f_i^\alpha(\tau;\bs{k})
+\sum_j\tilde{\calm}^2_{ij}(\tau)f_j^\alpha(\tau;\bs{k})=0\pkt
\end{equation}
The latin subscripts refer to the field components, $i=1,\dots,N$ 
and the greek superscripts $\alpha=1,\dots,N$
refer to the $N$ independent solutions of the system of coupled 
differential equations \eqn{eq:modeeq}. We will specify them
below, by a suitable set of initial conditions.

The Green's functions can be expressed by their Fourier components 
\begin{equation}
\tilde G_{ij}(\tau,\bs{x};\tau',\bs{x}')=\int\frac{d^{n-1}k}
{(2\pi)^{n-1}}e^{i\bs{k}(\bs{x}-\bs{x}')}\tilde G_{ij}(\tau,\tau';\bs{k})\pkt
\end{equation}
These Fourier components can be rewritten in terms of mode functions. They read
\begin{eqnarray}
\tilde G_{ij}(\tau,\tau';\bs{k})&=&\sum_{\alpha\beta}
W^{-1}_{\beta\alpha}\left[f_i^\alpha(\tau,k)f_j^{*\beta}(\tau',k)\theta(\tau-\tau')\right.\nonumber\\
&&\left.+f_i^\alpha(\tau',k)f_j^{*\beta}(\tau,k)\theta(\tau'-\tau)\right]\kma
\end{eqnarray}
where $W_{\alpha\beta}$ is the Wronskian matrix of the system of solutions:
\begin{equation}
W^{\alpha\beta}=
\left[\sum_j
f _j^{\alpha*}f_j^{\beta}{}'-f_j^{\alpha*}{}'f_j^\beta\right]
\pkt
\end{equation}
Therefore the expectation values of the fluctuation fields are given by the fluctuation integrals:
\be 
\calf_{ij}=<\tilde\psi_i\tilde\psi_j>= -i \int
\frac{d^{n-1}k}{(2\pi)^{n-1}}
\sum_{\alpha\beta}W^{-1}_{\beta\alpha}{\rm Re\;}(f_i^\alpha f_j^{\beta*})\pkt
\ee
The Wronskian matrix is determined by the initial conditions
of the system of fluctuations which will be discussed
in the next subsection.


\subsection{Initial conditions}
\label{subsec:initialconditions}

The choice of initial conditions for the quantum system is very 
important for the renormalisation of the equations of motion.
In order to satisfy the canonical commutation relations for the 
creation and annihilation operators and choose a Fock space basis, 
we first diagonalise the mass matrix $\tilde \calm^2_{ij}$ at 
the initial time. The eigenvalues $m^2_{\alpha0}$, $\alpha=1,N$ 
then define $N$ independent free particle Fock spaces whose product 
we choose as our initial state.
The eigenvalues and eigenvectors of the initial mass matrix 
$\tilde\calm^2_{ij}(0)$ are 
defined by the equations
\be
\det \left\{\tilde\calm^2_{ij}(0)-m_{\alpha0}^2\delta_{ij}\right\}=0
\ee
and
\begin{equation}\label{eq:eigenvalueequation}
-m^2_{\alpha0}f_{i0}^\alpha+\tilde\calm^2_{ij}(0)f_{j0}^\alpha=0\pkt
\end{equation}
The eigenvectors $f_{i0}^\alpha$ are orthogonal, we choose
them  real and normalised as
\be
\sum_l f^\alpha_{l0}f^\beta_{l0}=\delta_{\alpha\beta}\kma\hspace{7mm}
\sum_\alpha f^\alpha_{l0}f^\alpha_{m0}=\delta_{lm}
\pkt\ee 
We now specify the mode functions introduced in the previous
section by the initial conditions
\bea \label{eq:init1}
f_l^\alpha(0,k)&=& f_{l0}^\alpha\kma 
\\
\label{eq:init2}
{f_{l}^{\alpha}}'(0,k)&=&-i\Omega_{\alpha 0}f_{l0}^\alpha
\kma\eea 
with the frequencies $\Omega_{\alpha0}(0)=\sqrt{k^2+m_{\alpha0}^2}$. 
The Wronskian matrix of the mode functions at time $\tau=0$ is given
by
\bea \nonumber
&&W_{\alpha\beta}=W(f^\alpha(0,k),f^\beta(0,k))
\\&&=\sum_l \left[f_{l}^{\alpha*}(0,k) f_{l}^{\beta}{}'(0,k)
-f_{l}^{\alpha*}{}'(0,k)f_{l}^\beta(0,k)\right]
=-2i\Omega_{\alpha0}\delta_{\alpha\beta}\pkt
\eea
It is time-independent by construction. With these initial conditions 
the fluctuation integrals are given by
\be \label{eq:fluctuationintegrals}
\calf_{ij}=<\tilde\psi_i\tilde\psi_j>= \int
\frac{d^{n-1}k}{(2\pi)^{n-1}}\sum_\alpha
\frac{1}{2\Omega_{\alpha 0}}
{\rm Re \;}(f_i^\alpha f_j^{\alpha*})\pkt
\ee
These initial conditions correspond to a quantum state
which is denoted as the adiabatic vacuum. For such an initial state the 
energy-momentum tensor is found to be singular
at initial time \cite{Baacke:1997zz}. This problem can be 
solved by a suitable Bogoliubov transformations of the 
adiabatic vacuum. For a coupled system
this Bogoliubov transformation was derived in Ref. \cite{Baacke:2009sb}.
We recall the essential features of this approach in Appendix
\ref{app:initialsingularity}. As discussed in Ref. \cite{Baacke:1997zz}
this modification of the initial
state does not affect the renormalisation, which is our main subject here;
so we continue to work
with the adiabatic vacuum as the initial state.


\subsection{The energy-momentum tensor}
\label{subsec:energymomentumtensor}

In order to write the Einstein's equations we need the expression 
of the energy-momentum tensor for our system of coupled field 
in Eq.~\eqn{curvedlagrangian}.
Generalising straightforwardly the case of a single scalar field 
given in Ref.~\cite{Birrell:1982ix}, we have
\begin{eqnarray}
T_{\mu\nu}&=&\sum_i\left[(1-2\xi_i)
\Phi_{i;\mu}\Phi_{i;\nu}+
\left(2\xi_i-\frac{1}{2}\right)g_{\mu\nu}
g^{\rho\sigma}\Phi_{i;\rho}\Phi_{i;\sigma}
-2\xi_i\Phi_{i;\mu\nu}\Phi_i\nonumber\right.\\
&&\left.
+2\xi_ig_{\mu\nu}\Phi_i\Box\Phi_i
-\xi_i G_{\mu\nu}\Phi_i^2\right]
+g_{\mu\nu}V(\Phi)\pkt
\eea
In the Friedmann equations only $T_{tt}$ and $T_\mu^\mu$ appear
and their classical parts are given by
\begin{eqnarray}\label{eq:classicalenergy}
T_{tt}^{\rm cl}&=&\sum_i
\left(\frac{1}{2}\dot{\phi_i}^2-\xi_i G_{tt}\phi_i^2+
2(n-1)\xi_i H\phi_i\dot{\phi_i}\;
\right)+V^{cl}(\phi)
\; ,\\\nonumber
T_{\mu}^{\rm cl\;\mu}&=&
\sum_i\left\{\left[1-\frac n 2 +2(n-1)\xi_i\right]\dot{\phi_i}^2
+2(n-1)\xi_i\left[\ddot{\phi_i}+(n-1)H\dot{\phi_i}\right]\phi_i\right. 
\\\label{eq:classicaltrace}
&&\left.-\xi_i G_{\mu}^{\mu}\phi_i^2\right\}+n V^{cl}(\phi)
\kma\end{eqnarray}
with
\be
V^{\rm cl}(\phi)=\frac{1}{2}\sum_i m_i^2 \phi_i^2+ 
\frac{1}{4}\sum_{i,j}\lambda_{ij}\phi_i^2\phi_j^2
\pkt\ee
Using conformal time and the rescaled fields we 
find~\footnote{Though we have introduced conformal time
we continue to use  the time-time  component of the 
energy-momentum tensor in standard time $t$, as made explicit by 
using time indices $t$ instead of $0$ or $\tau$.}
\bea\nonumber 
T_{tt}^{\rm cl}&&=\frac{1}{a^n}\biggl\{
\frac{1}{2}\sum_i(\tilde\phi_i'^2+ m_i^2 a^2\tilde\phi_i^2)
\\\nonumber&&+
2(n-1)\sum_i\left(\xi_i-\xi_n\right)
\left(a H\tilde\phi_i'-\frac{n-2}{4}
a^2H^2\tilde\phi_i\right)\tilde\phi_i
\\\label{eq:tttrescaled}
&&+\frac{a^{4-n}}{4}\sum_{ij}\lambda_{ij}
\tilde\phi_i^2\tilde \phi_j^2
\biggr\}\pkt
\eea
For the fluctuation contribution to the energy density 
we obtain instead
\bea
&&T_{tt}^{\rm q}=\frac{1}{a^n}
\int\frac{d^{n-1}k}{(2\pi)^{n-1}}
\sum_\alpha \frac{1}{2\Omega_{\alpha0}}
\\\nonumber &&\times \left\{\frac{1}{2}\sum_i \left[|{f_i^{\alpha}}'|^2
+k^2|f_i^{\alpha}|^2\right]\right. 
+\frac{1}{2}\sum_{ij}
\tilde\calm^2_{ij}(\tau)f_i^{\alpha*}f_j^\alpha
\\ \nonumber
&&\left.\!\!+(n-1)\sum_i \left(\xi_i-\xi_n\right)\!\!
\left[aH\frac{d}{d\tau}|f_i^\alpha|^2\!-\!
\left(\frac{n-2}{2}a^2H^2+\frac{Ra^2}{2(n-1)}\right)|f_i^\alpha|^2
\right]\right\}\nonumber
\pkt\eea
The last term, proportional to $R$, which has no analogy in the 
classical energy-momentum tensor, is not genuine
but compensates an analogous term in  $\tilde\calm^2_{ij}$. 
It is convenient to rewrite $T_{tt}^{\rm q}$ as a function of
the fluctuation integral as
\bea \label{eq:tttelegant}
T_{tt}^{\rm q}&=&T_{tt}^{\rm q,kin}
+ \frac{1}{2a^n}\calv_{ij}\calf_{ij}
\\ \nonumber
&&+\frac{n-1}{a^n}\sum_i \left(\xi_i-\xi_n\right)
\left[aH\frac{d}{d\tau}\calf_{ii}-
\left(\frac{n-2}{2}a^2H^2+\frac{Ra^2}{2(n-1)}\right)\calf_{ii}
\right]\nonumber
\kma\eea
where we have introduced the "potential" 
\begin{equation}\label{eq:calV}
\calv_{ij}(\tau)=\tilde\calm^2_{ij}(\tau)-\tilde\calm^2_{ij}(0)\kma
\end{equation}
and where
\bea \nonumber
T_{tt}^{\rm q,kin}&=&\frac{1}{a^n}
\int\frac{d^{n-1}k}{(2\pi)^{n-1}}
\sum_\alpha \frac{1}{2\Omega_{\alpha0}}
\left\{\frac{1}{2}\sum_i \left[|{f_i^{\alpha}}'|^2
+\Omega_{\alpha0}^2|f_i^{\alpha}|^2\right]\right. 
\\&&\left.+\frac{1}{2}\sum_{ij}
(\tilde\calm^2_{ij}(0)-m_{\alpha0}^2\delta_{ij})f_i^{\alpha*}f_j^\alpha\right\}
\eea
is the kinetic or free-field part of the energy density.

The classical part of the trace of the energy-momentum tensor
becomes 
\bea\label{eq:tmumuclass}
T^{{\rm cl} \, \mu}_{\mu}&=&\frac{1}{ a^n}\left\{
2(n-1)\sum_i\left(\xi_i-\xi_n\right)\left[\tilde\phi_i'
-\frac 1 2 (n-2)aH\tilde\phi_i\right]^2\right.
\\\nonumber
&&\left.\!\!+2(n-1)\sum_i\xi_i\tilde\phi_i\tilde\phi_i''+
n\!\left[\frac 1 2\sum_i m_i^2\tilde\phi_i^2+
\frac{1}{4}\sum_{ij}\lambda_{ij}a^{4-n}\tilde\phi_i^2
\tilde\phi_j^2 + \Lambda\right]
\right\}
\pkt\eea
For the quantum contribution one finds instead
\bea\nonumber
T^{{\rm q} \,\mu}_{\mu}&=&\frac{1}{a^n}\int
\frac{d^{n-1}k}{(2\pi)^{n-1}}\sum_\alpha
\frac{1}{2 \Omega_{\alpha0}}\\&&
\times\left\{2(n-1)\sum_i(\xi_i-\xi_n)\left[ |{f_i^{\alpha}}'|^2
-k^2|f_i^\alpha|^2-\calm_{ij}^2f_i^{\alpha*}f_j^\alpha
\right.\right.\nonumber\\
&&\left.\left.-\frac{n-2}{2}a H  \frac{d}{d\tau}|f_i^\alpha|^2+
\frac{1}{4}\left((n-2)^2 H^2-\frac{n}{n-1}R\right)|f_i^\alpha|^2
\right]\nonumber\right.\\\label{eq:tmumuquantum}
&&\left.+\sum_{ij}\calm^2_{ij}f_i^{\alpha*}f_j^\alpha\right\}
\pkt\eea
Using the equation of motion for the mode functions,
Eq. \eqn{eq:modeeq}, the first three terms in the bracket 
can we rewritten as
\be
|{f_i^{\alpha}}'|^2
-k^2|f_i^\alpha|^2-\calm_{ij}^2f_i^{\alpha*}f_j^\alpha=
|{f_i^{\alpha}}'|^2+f_i^{\alpha^*}{f_i^{\alpha\,}}''=\frac{1}{2}
\frac{d^2}{d\tau^2}|f_i^{\alpha}|^2
\pkt\ee
We therefore may express the entire quantum contribution to
$T^\mu_\mu$ in terms of the fluctuations integral $\calf_{ij}$ as
\bea
T^{{\rm q} \,\mu}_{\mu}&=&\frac{n-1}{a^n}\sum_i (\xi_i-\xi_n)\left[
\frac{d^2}{d\tau^2}\calf_{ii}-(n-2)aH\frac{d}{d\tau}\calf_{ii}\right.
\\\nonumber
&&\left.
+\frac{1}{2}\left((n-2)^2 H^2-\frac{n}{n-1}R\right)\calf_{ii}\right] 
+\frac{1}{a^n}\sum_{ij}\calm_{ij}^2\calf_{ij}
\pkt
\nonumber
\label{eq:traceintermsoffij}
\eea


\section{ Renormalisation}
\label{sec:renormalization}
\setcounter{equation}{0}

In the previous sections we have introduced the fluctuation integrals
and expressed the equations of motion and the one-loop contributions 
to the energy-momentum tensor in term of these quantities.
These definitions have only formal character, as they involve divergent 
momentum integrals. As in the case of renormalisation in Minkowski, 
we have to introduce regulators in order to render the integrals 
well-defined, and counterterms to absorb the divergences when the 
regulators are removed.
We will use here dimensional regularisation with the space dimension 
$3-\epsilon$. Furthermore, it is convenient to separate the limit 
$\epsilon \to 0$ entirely from the numerical computations by
adding and subtracting appropriate analytical terms, which reproduce
the divergent behaviour and render the numerical integrations finite. 
Such a scheme was set up in Ref. \cite{Baacke:1996se}, and we
will use it here.  We recall its main features and the 
formulae needed here in Appendices \ref{app:perturbativeexpansion},
\ref{app:largemomentumbehavior}
and \ref{app:dimensionallyregulatedintegrals}.   

Before we discuss renormalisation we have to discuss a subtle
issue (see also Ref. \cite{Baacke:1999gc}).
We have written all equations for general $n=4-\epsilon$, and the
equations of motion ensure that the energy-momentum tensor is
conserved in $n$ dimensions. Ultimately we have to take the
limit $n\to 4$, or $\epsilon\to 0$ and this will generate terms
that behave as $1/(n-4)$ from the divergent integrals and
consequently in the counterterm action. 
On the other hand the equations of motion and the energy-momentum 
tensor are equal to their $4$-dimensional form only up to terms 
of order $(n-4)$. These terms will survive the limit $n \to 4$ 
whenever they are multiplied by a factor $1/(n-4)$.
This will result in ``additional finite terms'', which have no
counterpart in quantum field theory in Minkowski space.
So our prescription is: we first formulate the full renormalised 
theory in $n$ dimensions and take the limit $n\to 4$ only at the end.
Another prescription one may think to use, is consider $ n \neq 4 $ 
only in the divergent integrals, while keeping the equation of
motion and the energy-momentum tensor in 4 dimensions, but in that 
case we cannot rely on energy-momentum covariant conservation
during the whole regularisation procedure.


\subsection{Renormalisation of the equation of motion}
\label{sec:renormeqmot}
The equation of motion contains logarithmic and quadratic divergences
due to the fluctuation integrals. Using the analysis of Appendix
\ref{app:perturbativeexpansion} and in particular 
Eq.~\eqn{eq:logdivergence}, their divergent behaviour can be
understood and it coincides with that of the divergent integral
\begin{equation}
 \calf_{ij}^{div}=a^\epsilon\int\frac{d^{n-1}k}{(2\pi)^{n-1}}
\sum_\alpha\frac{1}{2\Omega_{\alpha0}}\left[f_{i0}^\alpha f_{j0}^\alpha
-\sum_\beta f_{i0}^\alpha f_{j0}^\beta
\frac{\tilde\calv_{\alpha\beta}(\tau)}{\Omega_{\beta0}
(\Omega_{\alpha0}+\Omega_{\beta0})}\right]\kma
\end{equation}
where the potential $\tilde \calv_{\alpha\beta}$ has been introduced
in Eq. \eqn{eq:tildecalV}
Here and in the following we use a slight change in notation,
without introducing a new symbol: we incorporate the 
prefactor $a^\epsilon$ which appears in front of the
divergent integrals into the $n$-dimensional
integration measure.

Using the formulae for the various integrals given in
Appendix \ref{app:dimensionallyregulatedintegrals} one can obtain
analytically the dimensionally regulated form of the previous 
expression as 
\begin{eqnarray}\label{calfreg}
\calf_{ij}^{\rm reg}&=&-\frac{L_\epsilon}{16\pi^2}
\tilde\calm_{ij}^2(\tau)
+\sum_\alpha\frac{m_{\alpha0}^2}{16\pi^2}
f_{i0}^\alpha f_{j0}^\alpha\left(\ln\frac{m_{\alpha0}^2}{a^2\mu^2}
-1\right)\nonumber\\
&&+\sum_{\alpha,\beta}f_{i0}^\alpha f_{j0}^\beta 
\frac{\tilde\calv_{\alpha\beta}(\tau)}{16\pi^2}
\left(\ln\frac{m_{\alpha0}^2}{a^2\mu^2}
-\frac{m_{\beta0}^2}{m_{\alpha0}^2
-m_{\beta0}^2}\ln\frac{m_{\beta0}^2}{m_{\alpha0}^2}-1\right)\kma
\end{eqnarray}
where $L_\epsilon=\frac{2}{\epsilon}-\gamma+\ln{4\pi}$.
In order to cancel this divergence in the equations of motion we have to
add an appropriate counterterm to the Lagrange density. In the case
of Minkowski space-time, it was found in Ref. \cite{Baacke:2003bt} that 
the counterterm Lagrangian 
\begin{equation}
  \call^{\rm ct}_{Mink.}=-\delta\zeta 
\calm^2_{kl} \calm_{lk}^2
\end{equation}
with 
\be
\delta \zeta=\frac{L_\epsilon}{64\pi^2}
\ee
takes account of all divergences. We work here in the $\overline{MS}$
scheme and subtract only the divergent part of the fluctuation integral.
Note that for constant masses and fields, this is a contribution to 
the renormalisation of the cosmological constant of order $ \sum_i m_i^4 $ 
and also contains mass and coupling renormalisation terms due to the 
interaction part of the effective mass matrix $ \calm_{ij}^2$. 

Taking into account the factor $\sqrt{-g}=a^n$ (using conformal time), 
and noting that the $\tilde \calm_{ij}^2$ differ by a factor
$a^2$ from $\calm_{ij}^2$, the equivalent Lagrangian for general
FRW is
\begin{equation}\label{eq:countertermlagrangian}
 \call^{{\rm ct}}=-a^{n-4}\delta\zeta \tilde
\calm^2_{kl}\tilde \calm_{lk}^2
\pkt\end{equation}
When discussing the energy-momentum tensor it will be convenient
to have this Lagrangian decomposed into the usual mass, coupling 
constant and other gravity-related counterterms. This is presented 
in Appendix \ref{sec:counterterms}.

Introducing this counterterm Lagrangian and given that
\bea\nonumber
\frac{\partial}{\partial \tilde \phi_i}\call^{\rm ct}
&=&2
\delta \zeta \tilde \calm^2_{kl}\left(4\lambda_{il}\tilde \phi_l \delta_{ik}
+2\delta_{kl}\lambda_{ki}\tilde\phi_i\right)\\
&=&4\delta\zeta\sum_l
\left[2\tilde\calm^2_{il}\lambda_{il}\tilde \phi_l
+\lambda_{li}\tilde\calm^2_{ll}\tilde\phi_i\right]
\kma \eea
the equations of motion for the classical fields become
\bea
&&\tilde\phi_i''+(m_i^2+(\xi_i-\frac{1}{6})R)a^2\tilde\phi_i\\
\nonumber
&&+\sum_j\lambda_{ij}\left[(\phi_j^2+4\delta\zeta
\tilde\calm^2_{jj}+\calf_{jj})\tilde\phi_i+(8\delta\zeta \tilde\calm_{ij}^2+
2\calf_{ij})\tilde\phi_j\right]=0\pkt
\eea
It follows from Eq. \eqn{calfreg} and the renormalisation procedure
that the combination $4\delta \zeta \tilde\calm_{ij}^2+\calf_{ij}$
is finite. This is entirely analogous to the case without
coupling to gravity. There is, however, one important difference:
in addition to the renormalisation scale the scale factor $a(\tau)$
appears in the prefactor $a^\epsilon$  and this modifies the
finite terms. In order to be able to identify the different
contributions to $ \calf_{ij}$ separately we write 
\be
\calf_{ij}= -\frac{L_\epsilon}{16\pi^2}
\tilde\calm_{ij}^2(\tau)+\calf_{ij}^{\rm fin}+\calf_{ij}^{\rm add}
\kma\ee
where the finite part of the fluctuation integral is given by
\be
\calf_{ij}^{\rm fin}=\calf_{ij}^{\rm sub} + \calf_{ij}^{\rm ft}
\pkt\ee
Here we define
\be\label{eq:calfijsub}
\calf_{ij}^{\rm sub}=
\int\frac{k^2 dk}{2\pi^2}\sum_\alpha\frac{1}{2\Omega_{\alpha0}}
\left[{\rm Re \;}(f_i^\alpha f_j^{\alpha*})-f_{i0}^\alpha f_{j0}^\alpha
+\sum_\beta f_{i0}^\alpha f_{j0}^\beta
\frac{\tilde\calv_{\alpha\beta}(\tau)}{\Omega_{\beta0}(\Omega_{\alpha0}+\Omega_{\beta0})}\right]
\ee 
as the {\em subtracted fluctuation integral}, that is finite, but
in  general has to be computed numerically.
From $\calf_{ij}^{\rm reg}$, Eq. \eqn{calfreg}, we have
\bea \label{eq:calfft}
\calf_{ij}^{\rm ft}&=&\sum_\alpha\left[\frac{m_{\alpha0}^2}{16\pi^2}f_{i0}^\alpha 
f_{j0}^\alpha\left(\ln\frac{m_{\alpha0}^2}{a^2\mu^2}-1\right)
\right.\nonumber\\
&&\left.+\sum_\beta f_{i0}^\alpha f_{j0}^\beta\frac{\tilde\calv_{\alpha\beta}(\tau)}{16\pi^2}
\left(\ln\frac{m_{\alpha0}^2}{a^2\mu^2}-\frac{m_{\beta0}^2}{m_{\alpha0}^2-m_{\beta0}^2}\ln\frac{m_{\beta0}^2}{m_{\alpha0}^2}-1\right)\right]
\eea
as {\em finite terms} that are left over after removing the divergent
part. 
However, among those we have identified separately
\be
\calf_{ij}^{\rm add}=-\frac{1}{288\pi^2}a^2 R \delta_{ij}
\kma\ee 
which arises from the $ {\cal O} (\epsilon) $ term in the expansion
of $ \tilde{\calm}^2_{ii} $:
\be
\tilde{\calm}^2_{ii}=
\left(m_i^2+(\xi_i-1/6)R\right)a^2
+a^{4-n}\left(3\lambda_{ii}\tilde{\phi_i}^2
+\sum_{j\neq i}\lambda_{ij}\tilde\phi_j^2\right)+\frac{\epsilon}{36}a^2 R
+ O(\epsilon^2)\pkt
\ee
This is the only term that has no usual $4D$ counterpart and is
generated from taking the $n$-dimensional FRW model. It gives
a finite contribution to the scalar field coupling to gravity $\xi_i$ 
and it may be in principle absorbed into a more general renormalisation
scheme than $\overline{MS} $. Note that such term survives even in the
case of 4D conformal coupling $ \xi_i = 1/6 $. 

Finally, the renormalised equation of motion for the
classical fields reads
\bea\label{eq:eqmclfren}
&&\tilde\phi_i''+(m_i^2+(\xi_i-\frac{1}{6})R)a^2\tilde\phi_i\\
\nonumber
&&+\sum_j\lambda_{ij}\left[(\tilde \phi_j^2+\calf^{\rm fin}_{jj}
+\calf_{jj}^{\rm add})\tilde\phi_i+2(\calf^{\rm fin}_{ij}
+\calf_{ij}^{\rm add})\tilde\phi_j\right]=0\pkt
\eea
Here we have taken the limit $\epsilon \to 0$ already,
and it is understood that the potential $\tilde\calv_{\alpha\beta}$
in $ \calf^{fin}_{ij} $ is computed in $4$ dimensions. Likewise, 
the  fluctuations are computed using the $4$-dimensional reduction
of $\tilde \calm_{ij}^2$ in their equation of motion.

Note that the occurrence of $\ln a(\tau)$ in the finite terms 
$\tilde \calm_{ij}^2$ is not necessarily a small correction. 
Neither is the ``potential'' $\calv_{\alpha\beta}$
small: the terms $(m_i^2+(\xi_i-1/6)R)$ appear multiplied by $a^2$.
Indeed the qualitative r\^ole of the  $\ln a(\tau)$ 
term in $\calf^{\rm ft}_{ij}$ is   to {\em compensate} a logarithmic increase 
of the subtracted integral $\calf_{ij}^{\rm sub}$.
This is illustrated by an example in Appendix \ref{app:anexample}.


\subsection{Renormalisation of the energy-momentum tensor}
\label{sec:energyrenorm}

We can now proceed in the same way to the renormalisation
of the energy-momentum tensor.
In Sec. \ref{subsec:energymomentumtensor}
we have separated the quantum part of the $tt$ component of the
energy-momentum tensor into a ``kinetic'' part and some further 
contributions which can be written in terms of the fluctuation
integrals. The divergences of the latter have been discussed in the
previous section, so it remains here to analyse the kinetic part.
Using the results of Appendix \ref{app:perturbativeexpansion}
and in particular Eq. \eqn{eq:kinetictermsimplified} 
we obtain
\begin{eqnarray}\nonumber
T_{tt}^{\rm q,kin}&=&
\frac{1}{a^n}\int\frac{d^{n-1}k}{(2\pi)^{n-1}}
\sum_\alpha
\frac{1}{2\Omega_{\alpha0}}\left[\Omega_{\alpha0}^2
+\frac{1}{2}|{h_i^{\alpha}}'|^2\right.
\\&&\left.+\frac{1}{2}(\tilde\calm_{ij}^2(0)-m_{\alpha0}^2\delta_{ij})
{\rm Re \;}(h_i^\alpha h_j^{\alpha*})\right]\pkt
\end{eqnarray}
The divergent behaviour of the first term in the bracket
is obvious, the one of the second term is given by Eq. 
\eqn{eq:kineticdivergence}. The third term originally
appears with the mode functions $f_i^\alpha$; the present form, with the 
reduce mode functions $h_i^\alpha$, is obtained by using that
the parenthesis $(\tilde\calm_{ij}^2(0)-m_{\alpha0}^2\delta_{ij})$
vanishes when multiplied with $f_{i0}^\alpha$ or $f_{j0}^\alpha$ due
to the eigenvalue equation \eqn{eq:eigenvalueequation}.
Still the term could lead to a divergence arising from the second order 
contribution $h^{(1)\alpha}_i h^{(1)\alpha*}_j$. One can convince oneself that
this leading term vanishes. The expansion of $h^{(1)\alpha}_i$
and   $h^{(1)\alpha*}_j$ contains factors
$f_{i0}^\beta$ or $f_{j0}^\beta$ (see Eqs.
\eqn{eq:reh1as} and \eqn{eq:imh1as}); the parenthesis then reduces to
$(m_{\beta0}^2-m_{\alpha0}^2)\delta_{ij}$ and this is found to be multiplied
with a term that is symmetric in $\alpha$ and $\beta$. 
So, using the integrals given in Appendix
\ref{app:dimensionallyregulatedintegrals} we can write
\be
T_{tt}^{\rm q,kin}= -\frac{1}{64 \pi^2 a^4}L_{\epsilon}
\left[\sum_\alpha m_{\alpha_0}^4- \sum_{\alpha\beta}\tilde \calv
_{\alpha\beta}\tilde \calv_{\beta\alpha}\right]
+T_{tt}^{\rm q,kin, fin}\pkt
\ee
Here
\be
T_{tt}^{\rm q, kin, fin}=
T_{tt}^{\rm q, kin, sub} + T_{tt}^{\rm q, kin, ft}\kma
\ee
where the subtracted integral is given by
\bea\nonumber
T_{tt}^{\rm q, kin, sub}&=&
\frac{1}{a^4}\int\frac{k^2dk}{2\pi^2}
\sum_\alpha
\frac{1}{2\Omega_{\alpha0}}\left[
\frac{1}{2}|{h_i^{\alpha}}'|^2-\sum_\beta 
\frac{\tilde\calv_{\beta\alpha}\tilde\calv_{\alpha\beta}}{4\Omega_{\beta 0}
(\Omega_{\alpha0}+\Omega_{\beta 0})}
\right.
\\&&\left.+\frac{1}{2}(\tilde\calm_{ij}^2(0)-m_{\alpha0}^2\delta_{ij})
{\rm Re \;}(h_i^\alpha h_j^{\alpha*})\right]
\label{eq:Tttkinsub}
\eea
and the finite terms of the regularised integrals are
\bea\nonumber
&&T_{tt}^{\rm q, kin, ft}=
\frac{1}{64 \pi^2 a^4}
\left[\sum_\alpha m_{\alpha_0}^4 \left(\ln \frac{m_{\alpha0}^2}{a^2\mu^2} 
-\frac{3}{2}\right)\right.\\&&\left. -
\sum_{\alpha\beta}\tilde \calv
_{\alpha\beta}\tilde \calv_{\beta\alpha}
\left(\ln\frac{m_{\alpha0}^2}{a^2\mu^2}
-\frac{m_{\beta0}^2}{m_{\alpha0}^2-
m_{\beta0}^2}\ln\frac{m_{\beta0}^2}{m_{\alpha0}^2}-1\right)
\right] \pkt
\eea
In both of these equations it is understood that $n=4$.

Besides $T_{tt}^{\rm q, kin}$ the $tt$ component of the energy-momentum
tensor contains further quantum contributions, which have been 
expressed by fluctuation integrals, see Eq. \eqn{eq:tttelegant}. 
Collecting all these finite pieces we define
\bea \label{eq:tttfinite}
&&T_{tt}^{\rm q, fin}=T_{tt}^{\rm q, kin, fin}
+ \frac{1}{2}\calv_{ij}\calf^{\rm fin}_{ij}
\\ 
&&+\frac{3}{a^4}\sum_i \left(\xi_i-\frac{1}{6}\right)
\left[aH\frac{d}{d\tau}\calf^{\rm fin}_{ii}-
\left(a^2H^2+\frac{Ra^2}{6}\right)\calf^{\rm fin}_{ii}
\right]\nonumber
\kma
\eea
where we have already set $n=4$. 

For the singular terms we find instead
\bea\nonumber
T_{tt}^{\rm q,sing}&=&- \frac{L_\epsilon}{64\pi^2}\frac{1}{a^n}
\tilde\calm_{ij}^2\tilde\calm_{ij}^2\\\nonumber
&&+\frac{n-1}{a^n}\sum_i(\xi_i-\xi_n)
\left(\frac{n-2}{2}a^2H^2+\frac{Ra^2}{2(n-1)}\right)
\frac{L_\epsilon}{16\pi^2}\tilde\calm_{ii}^2\\
&&-\frac{n-1}{a^n}\sum_i
(\xi_i-\xi_n)\frac{L_\epsilon}{16\pi^2}
\frac{d}{d\tau}\tilde\calm_{ii}^2 \pkt
\end{eqnarray} 
Here it is understood that everything is still defined for
general $n$.
The counterterms are (see Eqs. \eqn{eq:energymomentumtensorredef}
and \eqn{eq:tttrescaled})
\bea\nonumber
&&T_{tt}^{\rm ct}=\frac{1}{a^n}\biggl\{
\frac{1}{2}\sum_i\delta m_i^2 a^2\tilde\phi_i^2
\\\nonumber&&+
2(n-1)\sum_i \delta \xi_i
\left(a H\tilde\phi_i'-\frac{n-2}{4}
a^2H^2\tilde\phi_i\right)\tilde\phi_i
\\
&&+\frac{a^{4-n}}{4}\sum_{ij}\delta \lambda_{ij}
\tilde\phi_i^2\tilde \phi_j^2
\biggr\}
+\delta \tilde \Lambda + \delta \tilde Z G_{tt}+
\delta \tilde\alpha {}^{(1)}H_{tt}
\pkt
\eea
Adding the singular part and the counterterms one finds
that the divergences are cancelled, but, as in the case of
$\calf_{ij}$ some additional finite terms remain in the
limit $n\to 4$.
These are
\bea\nonumber
&&T_{tt}^{\rm q, add}=
\lim_{n\to 4} (T_{tt}^{\rm q, sing}+ T_{tt}^{\rm ct})
=\frac{H}{96\pi^2 a^3}
\sum_i\left(aH \tilde\calm_{ii}^2-\frac{d}{d\tau}\tilde\calm^2_{ii}\right)
\\
&&+ \frac{1}{16\pi^2}\sum_i\left(\xi_i-\frac{1}{6}\right)
\left[\frac{1}{36}{}^{(1)}H_{tt}+\frac{1}{72}R^2-6 \frac{H^2}{a^2}
\tilde \calm^2_{ii}\right]\pkt
\eea
It is understood that $n$ is set equal to $4$ in $\tilde\calm_{ii}^2$
and ${}^{(1)}H_{tt}$.
The result agrees with the one for the single-field case 
found in Ref. \cite{Baacke:1999gc}.
There $T_{tt}^{\rm q, add}$ was in addition presented in an expanded form,
using the explicit expressions for  $\tilde\calm_{ii}^2$
and ${}^{(1)}H_{tt}$. Here we refrain from displaying such a rather
lengthy formula. We should mention that the calculations for checking
the cancellation of divergences and for obtaining $T_{tt}^{\rm q, add}$
are quite cumbersome, they have been performed using the
computer algebra code REDUCE \cite{Hearn:2004}.
Note that these additional terms are time-dependent and similar 
time-dependent pieces are contained in the subtracted integrals 
eq.~(\ref{eq:Tttkinsub}), as it happens for the $\ln a(\tau)$ terms 
in $\tilde \calf_{ij} $. Both together ensure that the energy-momentum
tensor is covariantly conserved.

There is still a further finite contribution to the energy-momentum
tensor, the conformal anomaly, which can in any case not
be renormalised away \cite{Birrell:1982ix}.
As discussed there and in Ref. \cite{Baacke:1999gc} it cannot 
be determined within our framework but has to be taken over from a 
more general analysis. It is obtained by setting, for $N$ scalar fields, 
\be
\delta\tilde\beta=-\delta\tilde\gamma
=\frac{N}{2880\pi^2}L_\epsilon
\pkt\ee
As ${}^{(2)}H_{\mu\nu}=H_{\mu\nu}$ in $4$ dimensions, the $n=4$
part of these tensors cancels, and so does the singularity.
But there is a finite remainder from the extension to $n\neq 4$.
We obtain \cite{Baacke:1999gc}
\bea\nonumber
T_{tt}^{\rm ano}&=&\lim_{n \to 4}
\frac{N}{2880\pi^2}L_\epsilon\left({}^{(2)}H_{tt}-H_{tt}\right)
\\&=&\frac{N}{2880\pi^2}\left(H\frac{R'}{a}+R H^2+\frac{1}{12}R^2+3 H^4\right)
\pkt\eea
Notice that these additional terms are vanishing for the case
of pure de Sitter or radiation dominated expansion, so they are
negligible in the inflationary phase, but they could have an
effect in the reheating phase.

Finally, the renormalised $tt$ component of the energy-momentum tensor
is given by
\be
T^{\rm ren}_{tt}=T_{tt}^{\rm cl}+T_{tt}^{\rm q, fin}+T_{tt}^{\rm q, add}
+T_{tt}^{\rm ano}\pkt
\ee

As we have seen in Sec. \ref{subsec:energymomentumtensor},
 $T_\mu^{\rm{q}~\mu }$, the unrenormalised quantum contribution to
the trace of the energy-momentum tensor can entirely be expressed 
in terms of the fluctuation integrals
$\calf_{ij}$ and derivatives thereof, see Eq. \eqn{eq:traceintermsoffij}.
So the finite part is simply given by
\bea\nonumber
T^{{\rm q,fin} \,\mu}_{\mu}&=&\frac{3}{a^4}\sum_i (\xi_i-\frac{1}{6})\left[
\frac{d^2}{d\tau^2}\calf^{\rm fin}_{ii}
-2aH\frac{d}{d\tau}\calf^{ \rm fin}_{ii}\right.
\\
&&\left.
+\left(2 H^2-\frac{2}{3 }R\right)\calf_{ii}^{\rm fin}\right]
+\frac{1}{a^4}\sum_{ij}\tilde\calm_{ij}^2\calf^{\rm fin}_{ij}\pkt
\eea
Likewise, the divergences of $T_\mu^{{\rm q}~\mu}$ are related 
in a straightforward way to
those of the fluctuation integrals.
Therefore, the part that is singular as $n\to 4$, is given by
\bea\nonumber
T^{{\rm q,sing} \,\mu}_{\mu}&=&-\frac{L_\epsilon}{16\pi^2a^n}
\left\{(n-1)\sum_i (\xi_i-\xi_n)\left[
\frac{d^2}{d\tau^2}\tilde\calm^2_{ii}
-(n-2)aH\frac{d}{d\tau}\tilde \calm^2_{ii}\right. \right.
\\
&&\left.\left.
+\frac{1}{2}\left((n-2)^2 H^2-\frac{n}{n-1 }R\right)\tilde \calm^2_{ii}\right]
+\sum_{ij}\tilde \calm_{ij}^2\tilde\calm^2_{ji}\right\}\pkt
\eea
We have to add the appropriate counterterms which have the form
(see Eqs. \eqn{eq:energymomentumtensorredef} and \eqn{eq:tmumuclass})
\bea\label{eq:tmumucounterterm}
T^{{\rm q, ct} \, \mu}_{\mu}&=&\frac{1}{ a^n}\left\{
2(n-1)\sum_i\delta \xi_i\left[\tilde\phi_i'
-\frac 1 2 (n-2)aH\tilde\phi_i\right]^2\right.
\\\nonumber
&&+2(n-1)\sum_i\delta \xi_i\tilde\phi_i\tilde\phi_i''
\\\nonumber
&&\left.+
n\left[\frac 1 2\sum_i \delta m_i^2\tilde\phi_i^2+
\frac{1}{4}\sum_{ij}\delta \lambda_{ij}a^{4-n}\tilde\phi_i^2
\tilde\phi_j^2+\delta \tilde \Lambda\right]
\right\}
\\\nonumber
&&+\delta \tilde Z G_\mu^\mu + \delta\tilde \alpha {}^{(1)}H_\mu^\mu
\pkt\eea
We again find, using REDUCE, that the divergences cancel in the
sum of $T^{{\rm q, sing} \, \mu}_{\mu}$ and $T^{{\rm q, ct} \, \mu}_{\mu}$.
There remain finite terms, however, as for $T_{tt}$.
These are given by
\bea\nonumber
&&T^{{\rm q, add}\;\mu}_\mu=\lim_{n\to 4}(T^{{\rm q, sing} \, \mu}_{\mu}
+T^{{\rm q, ct} \, \mu}_{\mu})
=-\frac{1}{32\pi^2a^4}\sum_{i,j}\tilde\calm^2_{ij}\tilde\calm^2_{ji}
\\&&-\frac{1}{96\pi^2a^4}\sum_i\left[
\frac{d^2 \tilde \calm^2_{ii}}{d\tau^2}
-2aH\frac{d\tilde \calm^2_{ii}}{d\tau}
+2 a^2 H^2 \tilde\calm^2_{ii}\right]
\\\nonumber
&&\! -\!\frac{1}{16\pi^2}\!\sum_i\! \left(\xi_i-\frac{1}{6}\right)\!\!
\left[12\frac{H}{a^3} \frac{d\tilde\calm^2_{ii}}{d\tau}
\!+\!\frac{1}{a^2}(R-18 H^2)\tilde\calm^2_{ii}
\!-\!\frac{1}{18}R^2\!
-\!\frac{1}{36}{}^{(1)}H_\mu^\mu\right]\!\pkt
\eea
These agree, for the single-field case, i.e., $N=1$, with
Eq. (7.28) of Ref. \cite{Baacke:1999gc}, except for a misprint there:
$(R+18H^2)$ should read $(R-18H^2)$, as here. The expanded form $(7.29)$
in Ref. \cite{Baacke:1999gc} is correct. We again refrain from 
presenting such an expanded form.

The anomalous part is obtained as before
\bea \nonumber
T_\mu^{\rm ano\;\mu}&=&\lim_{n\to 4}
\frac{N}{2880\pi^2}L_\epsilon\left({}^{(2)}H_\mu^\mu-H_\mu^\mu\right)
\\&=&
\frac{N}{2880\pi^2}\left(\frac{R''}{a^2}+2 H \frac{R'}{a}+
2RH^2-12H^4\right)\pkt
\eea
Finally, we have to add the classical part, Eq. \eqn{eq:classicaltrace}, 
with $n=4$. So we finally have
\be
T_\mu^{{\rm ren}\;\mu}=
T_\mu^{{\rm cl}\;\mu}|_{n=4}+T_\mu^{{\rm q, fin}\;\mu}+
T_\mu^{{\rm q, add}\;\mu}+T_\mu^{\rm ano\;\mu}
\pkt\ee



\section{Simulations: the hybrid model}
\label{sec:simulations}
\setcounter{equation}{0}

While we have discussed here the renormalisation of a 
relatively general model of N coupled fields, the original incentive 
of our investigation was the application to the hybrid model 
of inflation.
In that case we have a system of only two fields and we can
write the Lagrangian as
\begin{eqnarray}\label{hybridlagrangian}
\call &=&\sqrt{-g}\left\{
\frac{1}{2} g^{\mu\nu}\partial_\mu\phi \partial_\nu\phi + 
\frac{1}{2} g^{\mu\nu}\partial_\mu\chi \partial_\nu\chi
- \frac{1}{2} m^2 \phi^2 - \frac{1}{4} \alpha (\chi^2 - v^2)^2
\right. \nonumber\\
& & \left.
- \frac{1}{2} \lambda \phi^2 \chi^2 
- \frac{1}{12} R \left( \phi^2 + \chi^2 \right)
\right\}\kma
\end{eqnarray}
Here $v$ is the vacuum expectation value of $\chi$. Furthermore,
we have to make use of a bare cosmological constant 
$\tilde \Lambda=\alpha v^4$ so that the minimum of the classical 
potential has the value zero.
Even when choosing the $\overline{MS}$ prescription for the
renormalisation of the masses and couplings we have to make sure that
the zero point of the quantum fluctuations at the minimum
vanishes, so we have to add a suitable finite 
counterterm $\delta \tilde\Lambda$. Of course this is ``fine tuning''
and represents the practical aspect of the cosmological constant
problem.
We have restricted our simulations to the case $\xi_1=\xi_2=1/6$,
the conformal couplings. This considerably simplifies the dynamics, 
and this choice would be natural in a supergravity theory.

\begin{figure}
	\centering
	\includegraphics[scale=0.7]{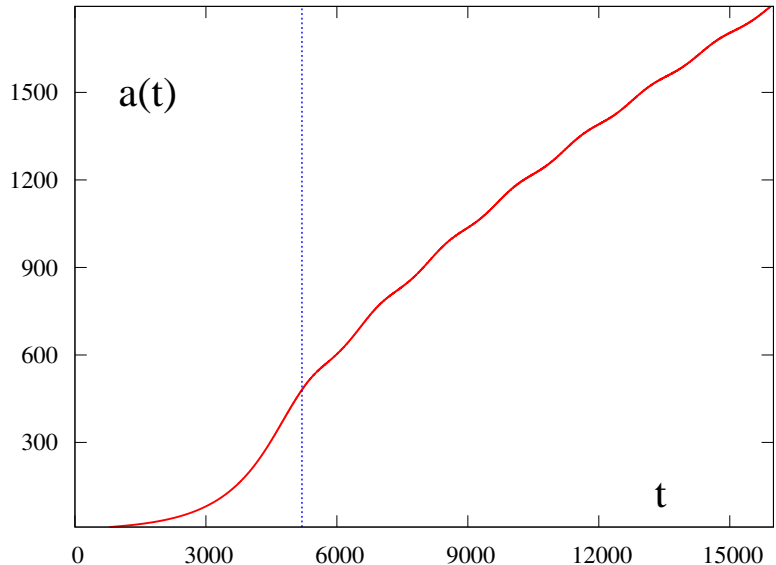}
	\caption{The evolution of the scale parameter $a(t)$ for set~I.}
	\label{fig:setI:a}
\end{figure}

\begin{figure}
	\centering
	\includegraphics[scale=0.7]{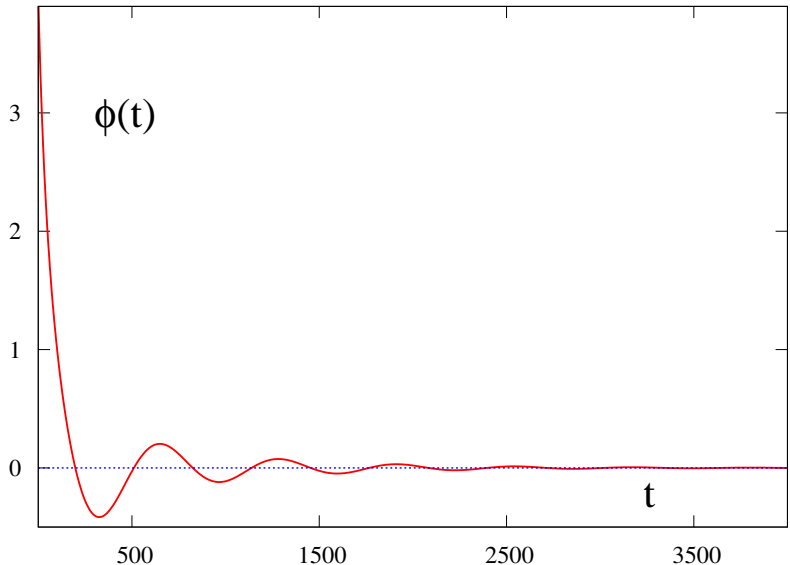}
	\caption{The evolution of the inflation field $\phi(t)$ for set~I.}
	\label{fig:setI:phi}
\end{figure}

\begin{figure}
	\centering
	\includegraphics[scale=0.7]{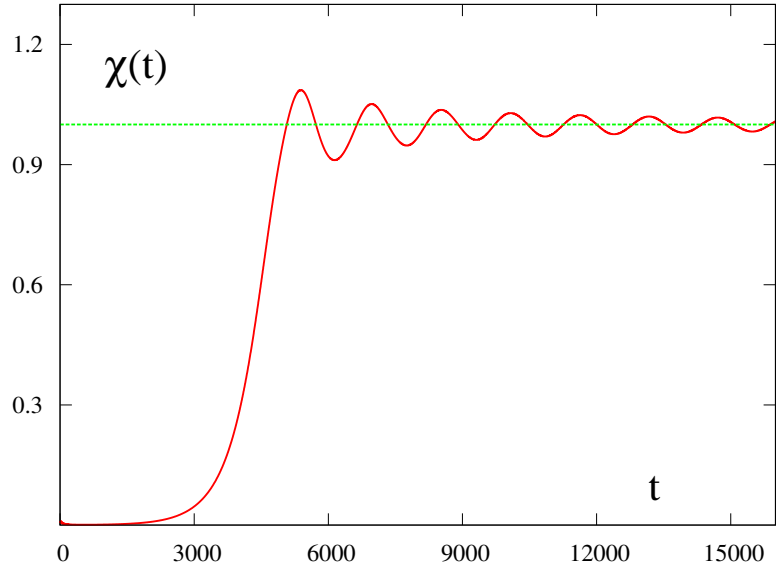}
\caption{The evolution of the waterfall field $\chi(t)$ for set~I.}
	\label{fig:setI:chi} 
\end{figure}

\begin{figure}
	\centering
        \includegraphics[scale=0.7]{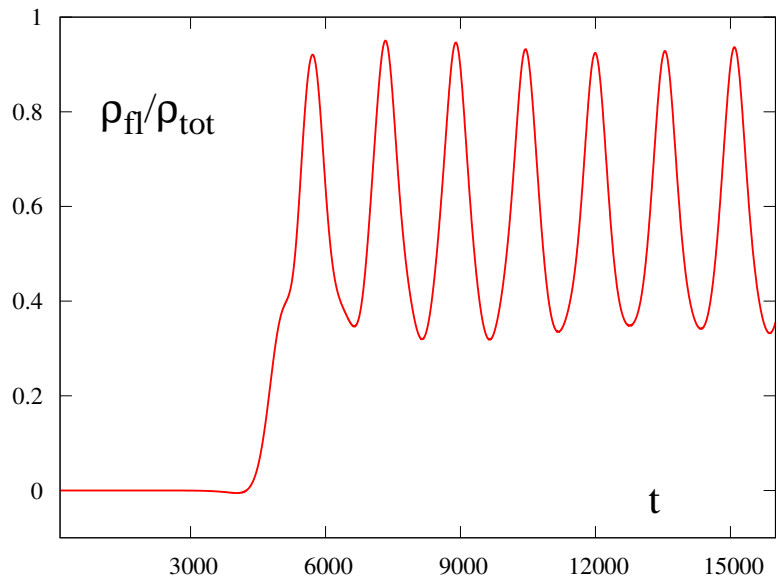}
		\caption{Ratio of fluctuation energy density to total 
energy density for set~I.}
	\label{fig:setI:qflqtot} 
\end{figure}

As parameter set $I$ we choose (in Planck units) $v=1$,
$m=0.01$, $\lambda=10^{-5}$, $\alpha=10^{-5}$.
The initial values of the fields are $\phi(0)=4$ and $\chi(0)=0.01$.
In Fig. \ref{fig:setI:a} we show the 
evolution of the scale parameter $a(t)$. We see an initial
exponential expansion, followed by a power behaviour $a \simeq t^{0.66}$
as expected for a matter dominated universe. The exponential behaviour
is not due to a slow roll of the ``inflaton'' field $ \phi$ which
oscillates with a rapidly decreasing amplitude, see Fig.
\ref{fig:setI:phi}. Rather it is the
field $\chi$, displayed in  Fig. \ref{fig:setI:chi} that 
takes some time before falling from the
metastable maximum to the stable minimum. Quantum fluctuations
build up after the field $\chi$ starts to fall from $\chi=0$ to
$\chi=v=1$. Of course both the classical as well as the
quantum energy density decrease with the expansion of the
universe. We therefore plot in Fig. \ref{fig:setI:qflqtot} 
the ratio $\rho^{\rm fl}/\rho$ of 
the fluctuation energy and the total energy densities. The fact that
the classical energy density is not converted entirely into fluctuation 
energy is typical for the one-loop approximation in which the
classical fields and the fluctuations remain coherently coupled.

The covariant energy conservation, Eq. \eqn{eq:covencons} with
$n=4$, is fulfilled with a relative error less than 1 ppm. 

\begin{figure}
\vspace*{110mm}
	\centering
        \includegraphics[scale=0.75]{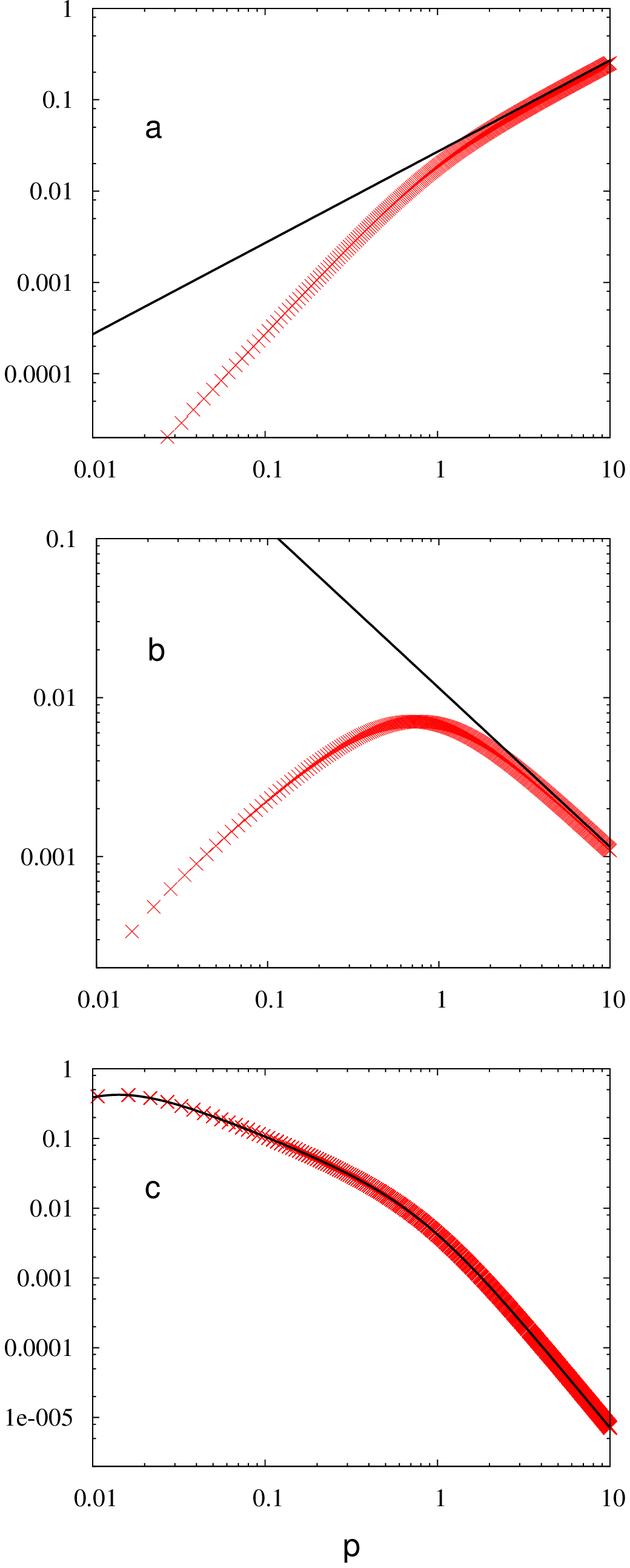}
		\caption{The integrand of the fluctuation integral
$\calf_{11}$: (a) without subtraction; (b) after removing the leading term;
and (c) with zeroth and first order in $\calv$ subtracted, respectively,
for set I; crosses: the numerical results; solid lines:
in parts a and b these indicate the asymptotic 
behaviour $p$ and $p^{-1}$, respectively;
in part c the solid line reproduces the approximation 
Eq.~\eqn{eq:eikonalapproximation}.}
	\label{fig:setI:F11} 
\end{figure}

\begin{figure}
  \centering
 \includegraphics[scale=0.7]{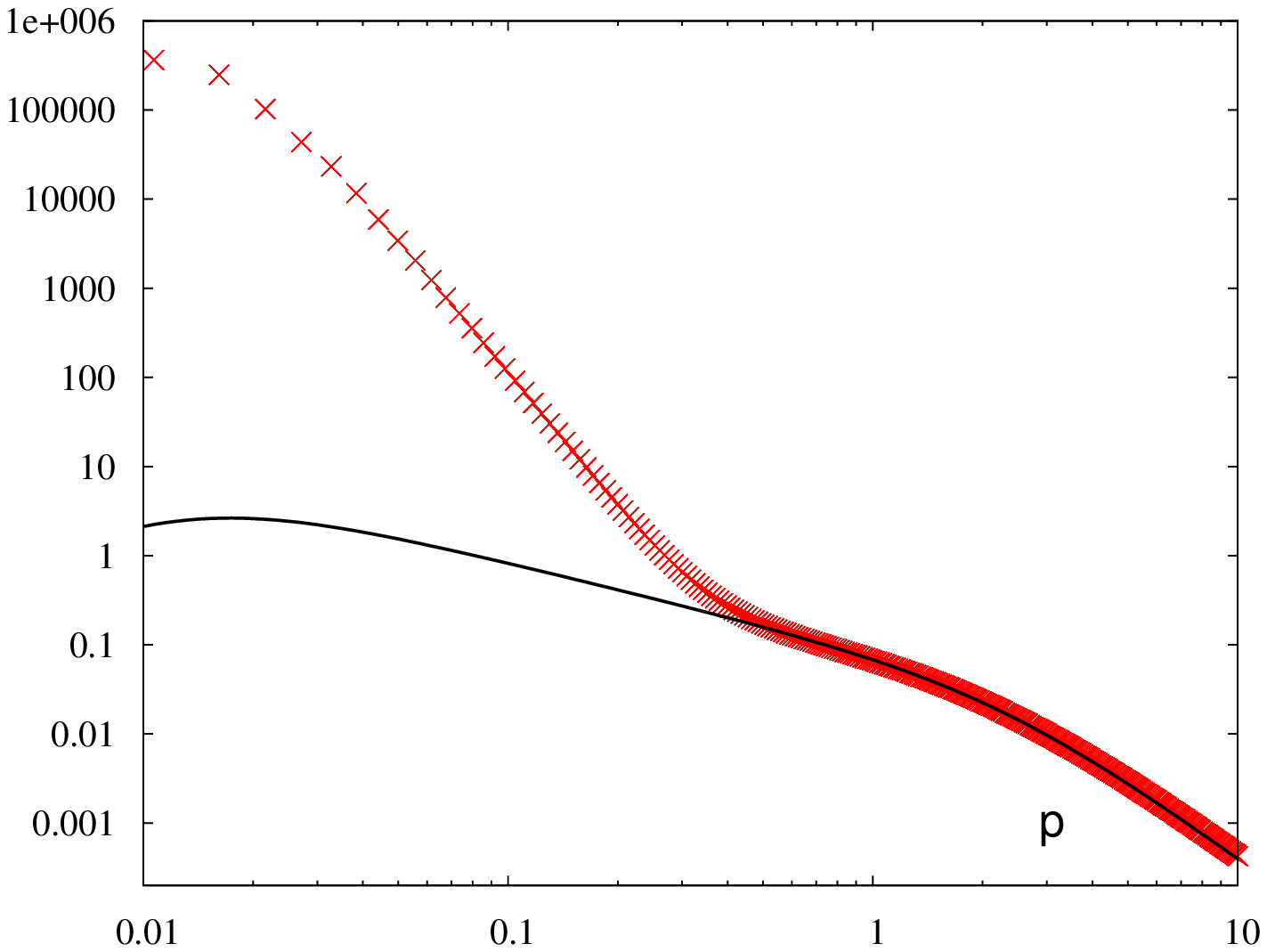}		
\caption{The integrand of the fluctuation integral $\calf_{22}$ for 
set I at \mbox{$t=6306$}; the crosses are the numerical results,
while the solid line is the asymptotic approximation 
Eq.~\eqn{eq:eikonalapproximation}.}
	\label{fig:setI:F22} 
\end{figure}

We also display, in Figs. \ref{fig:setI:F11}
and \ref{fig:setI:F22}, the integrands of the fluctuation integrals
$\calf_{11}$ and $\calf_{22}$, including the factor
$k^2/(2\pi^2)$. For $\calf_{11}$ we display the unsubtracted integrand,
the integrand with the zeroth order term removed, and fully subtracted
with zeroth and first order in $\calv$ removed, see Eq. \eqn{eq:calfijsub}.
The subtraction is actually performed using the functions $h_i^\alpha$ 
introduced in Appendix \ref{app:perturbativeexpansion} in order to
avoid small numerical differences. At large momenta the integrands 
are seen to behave as $p$, $p^{-1}$ and $p^{-3}$ respectively.
For the fully subtracted integrand we compare the numerical result
with the approximate formula \eqn{eq:eikonalapproximation}.
The agreement is almost perfect: the time at which this integrand is
extracted is $t=3154$. There the field $\phi$ has reached its asymptotic
value already and corrections due to the field $\chi$ are small.
The integrand for $\calf_{22}$ is plotted at $t=6306$. There the
field $\chi$ has come close to its asymptotic value $1$, but
is still oscillating around it. Here the approximate formula
\eqn{eq:eikonalapproximation} describes well the asymptotic
behaviour, but one notices a strong peak at low momenta
$p \lesssim .1$. This is in agreement with the fact that a significant
part of the classical energy has been converted into fluctuation energy,
as already displayed in Fig. \ref{fig:setI:qflqtot}. Obviously
it is the waterfall field whose fluctuations at low momenta 
have been strongly excited.  

\section{Conclusions}
\label{sec:conclusions}
\setcounter{equation}{0}

We have presented here a general analysis of the renormalisation 
of a set of coupled scalar fields in a spatially flat FRW universe 
with non-minimal gravitational coupling and higher curvature gravity
terms in the one-loop approximation. By a suitable subtraction 
procedure we obtain the equations of motion for the classical fields
and the energy-momentum tensor in renormalised form. The latter appears 
of course on the right-hand side of the Friedmann equations.
The counterterms are found in the standard form of quantum field
theory, they do not depend on numerical cutoffs nor on the initial 
conditions. The expressions to be evaluated on the computer are 
finite from the outset, in particular all the subtracted integrals
are finite and numerically well-behaved. 

The analysis of the divergences used here differs from the procedure
used often in the context of quantum field theory in curved space:
the adiabatic subtraction, using the eikonal expansion. 
We have not seen this approach being applied to a coupled channel 
system; as far as we see this meets problems of principle: while 
ordinary functions and their derivatives commute, this does not 
hold for matrix valued functions. Our renormalisation procedure
does not suffer from these shortcoming and can be applied to
the general case of coupled scalar fields. We expect it to be
easily generalised to contain fermionic fields as well and
be viable also for supersymmetric models.

The example of coupled scalar fields that has attracted most
attention is the hybrid model of inflation. We have now
implemented this renormalisation procedure for this specific
case and obtained first encouraging results: the energy-momentum
conservation is recovered in our scheme to high precision and 
our asymptotic formulae for the divergences describe the 
divergent fluctuation integrals with very good accuracy. 
In the renormalisation procedure we obtain additional finite 
terms,  some of which arise from the $ n $ dimensional FRW 
setting and have no 4D counterpart. These terms are in some
cases large and time-dependent, containing $ \ln (a(\tau)) $, 
but they cancel corresponding time-dependent terms in the 
fluctuation integrals and they are therefore crucial for 
obtaining consistent results, as is discussed in 
Appendix~\ref{app:anexample}. 

We have shown for a very simple case that our renormalisation
procedure is solid and gives very stable numerical results.
We plan to extend our study to the inflationary phase and
to different hybrid models in a future publication.


\section*{Acknowledgements}

One of us (N.K.) thanks the Humboldt Foundation for financial support, 
and the Deutsche Elektronensynchrotron DESY, Hamburg, for hospitality.
LC acknowledges the support of the DFG under the Collaborative Research 
Centre 676.


\appendix


\section{Tensors in n-dimensional flat FRW \\
geometry}
\setcounter{equation}{0}
\label{app:ndimtensors}

The extension of the FRW geometry to $n$ dimensions is done
in a specific way: the three-dimensional space of FRW
geometry is conformal to three-dimensional Minkowski space;
it is this three-dimensional space that we extend to
$3-\epsilon=n-1$ dimensions in the same way as it is done
in the usual dimensional continuation.

For the Christoffel symbols one finds
\bea
\Gamma\,^{t}_{ij}=a\dot{a}\,\delta_{ij}&\mbox{ ; }&
\Gamma\,^{j}_{kt}=\Gamma\,^{j}_{tk}=\frac{\dot{a}}{a}
\,\delta^j_k\pkt
\eea
The non-vanishing components of the Riemann tensor are 
\bea
R\,^t_{itk}=\ddot{a}a\,\delta_{ik}&\mbox{ , }&
R\,^t_{ijt}=-\ddot{a}a\,\delta_{ij}\kma\\
R\,^l_{ttk}=\frac{\ddot{a}}{a}\,\delta^l_k&
\mbox{ , }&R\,^l_{tjt}=-\frac{\ddot a}{a}\,\delta^l_j\kma
\eea
for those of the Ricci tensor one finds 
\bea
R\,_{tt}=(n-1)\frac{\ddot{a}}{a}&\mbox{ , }&R\,_{ij}=\left[-\ddot{a}a-(n-2)
\dot{a}^2\right]\delta_{ij}\pkt
\eea
This leads to the Ricci curvature scalar
\be
R=2(n-1)\frac{\ddot{a}}{a}+(n-1)(n-2)\left(\frac{\dot{a}}{a}\right)^2\pkt
\ee
Expressed in terms of Hubble's constant
\begin{equation} 
H(t)=\frac{\dot{a}(t)}{a(t)}\; 
\end{equation}
it takes the form
\be
R=(n-1)\left(2\dot{H}+nH^2\right)\pkt
\ee
The relevant components of the Einstein tensor become
\bea
G_{tt}&=&-\frac{(n-1)(n-2)}{2}H^2 \kma
\\
G_\mu^\mu&=&-\frac{n-2}{2}R
\pkt\eea

The time-time components and the trace of the tensors 
$\,^{(n)}H_{\mu\nu}$ are given by
\bea \label{htt1}
^{(1)}H\,_{tt}&=&-6H\dot R +\frac 1 2 R^2- 6 H^2R
\\ \nonumber &&+ (n-4)
\left(-2H\dot R-(n+1)RH^2\right)\kma\\ 
\label{htt2}
^{(2)}H\,_{tt}&=&-2H\dot R+\frac 1 6 R^2-2H^2R+(n-4)\left(-\frac 1 2 H\dot R
\right.\\ \nonumber
&&\left.-\frac{R^2}{24(n-1)}
-\frac 1 4(n+2)H^2R+\frac 1 8 (n-1)(n-2)^2H^4\right)\kma\\
\label{htt3}
H\,_{tt}&=&-2H\dot R+\frac 1 6 R^2-2H^2R\\ \nonumber
&&+(n-4)\left(
-\frac{R^2}{6(n-1)}-H^2R+\frac 1 2(n-1)(n-2)H^4
\right)\kma
\eea
\bea \label{hmumu1}
^{(1)}H\,_{\mu}^{\mu}&=&-6\ddot R-18H\dot R\\
&&+(n-4)\left(-2\ddot R-2(n+2)H\dot R-\frac 1 2 R^2\right)\kma\nonumber\\
\label{hmumu2}
^{(2)}H\,_{\mu}^{\mu}&=&-2\ddot R -6 H\dot R +(n-4)
\left(-\frac 1 2 \ddot R-\frac 1 2 (n+3)H\dot R
\right.\\ \nonumber
&&\left.-\frac{nR^2}{8(n-1)}+\frac 1 4 
(n-2)^2H^2R-\frac 1 8n(n-1)(n-2)^2H^4\right)\kma\\
\label{hmumu3}
H\,_{\mu}^{\mu}&=&-2\ddot R -6H\dot R +(n-4)
\left(-2H\dot R -\frac{R^2}{2(n-1)}
\right.\\&&\left.+(n-2)H^2R-\frac 1 2 n (n-1)(n-2)H^4\right)\nonumber
\pkt
\eea


\section{Removing the initial singularity}
\setcounter{equation}{0}
\label{app:initialsingularity}
The initial conditions we have discussed in Sec. \ref{subsec:initialconditions} 
lead after renormalisation to 
singularities in the fluctuation integrals when $\tau\rightarrow0$. They contain 
terms proportional to $\dot\calv_{\alpha\beta}(0)$ and $\ddot\calv_{\alpha\beta}(0)$
which behave as $\tau\ln[(m_{\alpha0}+m_{\beta0})\tau]$ and
$\tau^2\ln[(m_{\alpha0}+m_{\beta0})\tau]$ respectively. 
This does not affect the equations of motion, but the energy-momentum 
tensor, which contains the first and the second derivatives of the 
fluctuation integrals, that are infinite as time goes to zero. 
This means that these initial singularities will appear in the
Friedmann equations as well. In Ref. \cite{Baacke:2009sb} it was shown 
how to remove such singularities by a modification of the initial state. 
Here we will just briefly repeat these results.

The general concept of the Bogoliubov transformation supposes the replacement
of the naive initial state by a transformed vacuum state, annihilated by a 
superposition of annihilation $a_\alpha({\bf k})$ and creation $a^+_\alpha(-{\bf k})$
operators. For this purpose first we define the transformation
\begin{equation}
\tilde{a}_\alpha({\bf k})=\sum_\beta\sqrt{\frac{\Omega_{\alpha0}}{\Omega_{\beta0}}}
\left[C^{\alpha\beta}a_\beta({\bf k})-S^{\alpha\beta}a_\beta^+(-{\bf k})\right]
\end{equation}
and then a new vacuum state $|\tilde{0}>$
\begin{equation}
\tilde{a}_\gamma({\bf k})|\tilde{0}>=\sqrt{\frac{\Omega_{\gamma0}}{\Omega_{\alpha0}}}
C^{\gamma\alpha}\left[a_\alpha({\bf k})
-\sqrt{\frac{\Omega_{\alpha0}}{\Omega_{\beta0}}}
\rho_{\alpha\beta}a_\beta^+(-{\bf k})\right]|\tilde{0}>=0\kma
\end{equation}
with $\rho=C^{-1}S$ symmetric matrix.
By means of the expectation value of $a_\alpha({\bf k})a_\beta^+({\bf k}')$ in
the new vacua we define a matrix $M_{\alpha\beta}$
\begin{equation}
<\tilde{0}|a_\alpha({\bf k})a_\beta^+({\bf k}')|\tilde{0}>=(2\pi)^2\delta^3
({\bf k}-{\bf k}')2\sqrt{\Omega_{\alpha0}\Omega_{\beta0}}M_{\alpha\beta}(k)\kma
\end{equation}
which is related to the matrix $\rho$ as follows
\begin{equation}\label{eq:mrho}
M-\rho M^T\rho^+=I\pkt
\end{equation}
Following the schema presented in Appendix A of \cite{Baacke:2009sb} one sees that
the cancellation of the dangerous terms we have mentioned above requires
\begin{eqnarray}
\rm{Im} \rho^{\alpha\beta}&=&\frac{1}{2\Omega_{\alpha 0}\Omega_{\beta 0}}
\frac{1}{(\Omega_{\alpha0}+\Omega_{\beta0})^2}\calv'_{\alpha\beta}(0)\kma\\
{\rm Re\,} \rho^{\alpha\beta}&=&\frac{1}{2\Omega_{\alpha 0}\Omega_{\beta 0}}
\frac{1}{(\Omega_{\alpha0}+\Omega_{\beta0})^3}\calv''_{\alpha\beta}(0)\pkt
\end{eqnarray}

So after $\rho$ is known we have to determine the fluctuation integrals 
and the energy-momentum tensor in the Bogoliubov-transformed vacuum. 
But for the numerical implementation it is more convenient to redefine 
the mode functions 
\begin{equation}
\tilde f_i^\alpha(\tau;\bs{k})=
\sum_\beta\sqrt{\frac{\Omega_{\alpha0}}{\Omega_{\beta0}}}
\left[f_i^\beta(\tau;\bs{k})N_{\beta\alpha}
+f_i^{\beta*}(\tau;\bs{k})\rho^{\beta\gamma*}N_{\gamma\alpha}\right]\kma
\end{equation}
where the matrix $N$ is defined as $N\times N=M$. 
Here $\tilde f_i^\alpha$ are solutions
of the same mode equations as $f_i^\alpha$ are. For the fluctuation 
integrals and the energy-momentum tensor the expressions
presented in Sec.~\ref{subsec:energymomentumtensor}
remain valid, with the mode functions $f_i^\alpha$ 
replaced by the Bogoliubov
transformed ones~$\tilde f_i^\alpha$.


\section{Perturbative Expansion}

\label{app:perturbativeexpansion}

\setcounter{equation}{0}
The perturbative expansion of the mode functions for the case of 
coupled system in curved space-time follows a scheme
introduced in Ref. \cite{Baacke:1996se}.
We split the mode functions $f_i^\alpha(\tau;k)$ into the 
free part and higher order terms represented by the 
{\em reduced mode functions} $h_i^\alpha(\tau;k)$
with the ansatz
\begin{equation}
f_i^\alpha(\tau;k)=e^{-i\Omega_{\alpha 0}\tau}
[f_{i 0}^\alpha+h_i^\alpha(\tau;k)]\pkt
\end{equation}
The mode equations \eqn{eq:modeeq} are equivalent to the following integral equation 
\begin{equation}
f_i^\alpha(\tau;k)=e^{-i\Omega_{\alpha 0}\tau}f_{i0}^\alpha+\int_0^\tau d\tau'\Delta_{ij}(\tau-\tau';k)\calv_{jk}(\tau')f_k^\alpha(\tau';k)\kma
\end{equation}
with the potential $\calv_{ij}$ introduced in Eq. \eqn{eq:calV}
and with the retarded kernel of the free equation 
\begin{equation}
\Delta_{ij}(\tau-\tau';k)=\sum_\beta\frac{i}{2\Omega_{\beta 0}}\Theta(\tau-\tau')f_{i0}^\beta f_{j0}^\beta
[e^{i\Omega_{\beta 0}(\tau-\tau')}-e^{-i\Omega_{\beta 0}(\tau-\tau')}]\pkt
\end{equation}
Then the reduced mode functions satisfy the integral equation
\begin{eqnarray}
h_i^\alpha(\tau;k)&=&\int_0^\tau d\tau'\sum_\beta\frac{i}{2\Omega_{\beta 0}}f_{i0}^\beta f_{j0}^\beta
[e^{i(\Omega_{\alpha 0}+\Omega_{\beta 0})(\tau-\tau')}-e^{i(\Omega_{\alpha 0}-\Omega_{\beta 0})(\tau-\tau')}]\nonumber\\
&&\times\calv_{jk}(\tau')[f_{k0}^\alpha+h_k^\alpha(\tau';k)]
\end{eqnarray}
and the corresponding differential equations become
\begin{equation}
{h_i^{\alpha}}''-2i\Omega_{\alpha0}{h_i^{\alpha}}'(\tau;k)=-\sum_j
\left[\calv_{ij}f_{j0}^\alpha
+(\tilde\calm_{ij}^2-m_{\alpha0}^2\delta_{ij})h_j^\alpha(\tau;k)\right]\pkt
\end{equation}
The initial conditions \eqn{eq:init1} and \eqn{eq:init2}
imply $h_i^\alpha(0;k)={h_i^{\alpha}}'(0;k)=0$.

Now we expand the reduced mode functions 
$h_i^\alpha$ with respect to orders in potential $\calv_{ij}$
\begin{equation}
h_i^\alpha(\tau)=h_i^{\alpha(1)}(\tau)
+h_i^{\alpha(2)}(\tau)+h_i^{\alpha(3)}(\tau)+\ldots\kma
\end{equation}
where $h_i^{\alpha(n)}$ denotes 
the $n$th order in $\calv_{ij}$. For our 
calculations we will need only the first
and second terms of this expansion, therefore we 
introduce notation $h_i^{\alpha\overline{(n)}}(\tau)$ for the sum 
over all orders in $\calv_{ij}$ starting 
with the $n$th order, 
i.e. $h_i^{\alpha\overline{(1)}}=h_i^{\alpha(1)}+h_i^{\alpha\overline{(2)}}$.
Using the iteration one can obtain reduced mode functions of any order. 
In first order we have
\begin{eqnarray} 
h_i^{\alpha(1)}(\tau;k)&=&\sum_\beta\frac{i}{2\Omega_{\beta0}}f_{i0}^\beta f_{j0}^\beta f_{k0}^\alpha
\left[\int_0^\tau d\tau'\calv_{jk}(\tau')
e^{i(\Omega_{\alpha0}+\Omega_{\beta0})(\tau-\tau')}\right.\nonumber
\\\label{eq:h1}
&&\left.-\int_0^\tau d\tau'\calv_{jk}(\tau')
e^{i(\Omega_{\alpha0}-\Omega_{\beta0})(\tau-\tau')}\right]\pkt
\end{eqnarray}
The large-momentum behaviour of the Fourier-type integrals
is analysed in Appendix \ref{app:largemomentumbehavior}.
To leading order in $1/k$ we find
\be \label{eq:reh1as}
{\rm Re\;} h_i^{\alpha(1)}(\tau;k)
\simeq \sum_\beta
\frac{f_{i0}^\beta \tilde \calv_{\beta\alpha}(\tau)}{2 \Omega_{\beta0}(\Omega_{\alpha0}
+\Omega_{\beta_0})}
\ee 
and 
\be\label{eq:imh1as}
{\rm Im\;} h_i^{\alpha(1)}(\tau;k)\simeq
-\sum_\beta\frac{1}{2\Omega_{\beta0}}\int_0^\tau d\tau'
f_{i0}^\beta \tilde\calv_{\beta\alpha}(\tau')
\kma\ee
where we have introduced the notation
\be \label{eq:tildecalV}
f_{j0}^\beta f_{k0}^\alpha \calv_{ij}=
\tilde \calv_{\beta\alpha} 
\pkt
\ee
$\tilde \calv_{\alpha\beta}$ is symmetric in the indices, as
is $\calv_{jk}$.

For the fluctuation integrals \eqn{eq:fluctuationintegrals} we need to compute 
\begin{eqnarray}
{\rm Re \;}(f_i^\alpha f_j^{\alpha*})
&=&f_{i0}^\alpha f_{j0}^\alpha+f_{i0}^\alpha {\rm Re \;} h_j^\alpha
+f_{j0}^\alpha {\rm Re \;} h_i^\alpha
+{\rm Re \;}(h_i^\alpha h_j^{\alpha*})\nonumber\\
&=&f_{i0}^\alpha f_{j0}^\alpha
+f_{i0}^\alpha {\rm Re \;} h_j^{\alpha(1)}
+f_{j0}^\alpha {\rm Re \;} h_i^{\alpha(1)}
+f_{i0}^\alpha {\rm Re \;} h_j^{\alpha(\bar{2})}\nonumber\\
&&+f_{j0}^\alpha {\rm Re \;} h_i^{\alpha(\bar{2})}
+{\rm Re \;}(h_i^{\alpha(\bar{1})} h_j^{\alpha(\bar{1})*})\pkt
\end{eqnarray}
For large momenta the first term is constant, of course. The
second and third terms behave as $1/k^2$, see Eq. \eqn{eq:reh1as}.
The term ${\rm Re \;}(h_i^{\alpha(\bar{1})} h_j^{\alpha(\bar{1})*})$
obviously contains ${\rm Im \,}h_i^{\alpha(1)} {\rm Im \,}
h_j^{\alpha(1)}$ 
which likewise behaves as $1/k^2$, see Eq. \eqn{eq:imh1as}. It would
imply a {\em nonlocal} divergence.  This
contribution can be shown to cancel against a similar term
in $f_{i0}^\alpha {\rm Re \;} h_j^{\alpha(2)}
+f_{j0}^\alpha {\rm Re \;} h_i^{\alpha(2)}$, however.
For the single field case this has been discussed in Ref. \cite{Baacke:1996se}
(see there below Eq. (56)). 

The second and third term appear in
the integrand of fluctuation integral as:
\begin{eqnarray} \label{eq:logdivergence}
&&\sum_\alpha\frac{1}{2\Omega_{\alpha0}}
\left[f_{i0}^\alpha{\rm Re \;}h_j^{\alpha(1)}+f_{j0}^\alpha{\rm Re \;}h_i^{\alpha(1)}\right]\nonumber\\
&&\simeq
-\sum_{\alpha,\beta}f_{i0}^\alpha f_{j0}^\beta 
\frac{\tilde\calv_{\beta\alpha}}{2\Omega_{\alpha0}\Omega_{\beta0}
(\Omega_{\alpha0}+\Omega_{\beta0})}
\kma
\eea
where we have used Eq. \eqn{eq:reh1as}.

We have already used the Wronski relation at $\tau=0$ when 
defining the initial conditions in Sec. \ref{subsec:initialconditions}. 
Now we will consider the 
Wronskian in general, which will be useful for simplifying the 
mode integrals occurring in the energy-momentum tensor
\begin{equation}
W=f_i^{\alpha*}f_i^\beta{}'-f_i^{\alpha*}{}'f_i^\beta
=-2i\Omega_{\alpha0}\delta_{\alpha\beta}\pkt
\end{equation}
From this we obtain the relation
\begin{equation}
 i(\Omega_{\alpha0}+\Omega_{\beta0})(f_{i0}^\alpha h_i^\beta+f_{i0}^\beta h_i^{\alpha*}+h_i^{\alpha*}h_i^\beta)=f_{i0}^\alpha h_i^{\beta\prime}
-f_{i0}^\beta h_i^{\alpha*\prime}+h_i^{\alpha*}h_i^{\beta\prime}-h_i^{\alpha*\prime}h_i^\beta\pkt
\end{equation}
For our purposes we will need only the 
$\alpha=\beta$ part of this relation, i.e.:
\begin{equation}\label{eq:wronres}
\Omega_{\alpha0}(2f_i^\alpha{\rm Re \;}h_i^\alpha+|h_i^\alpha|^2)=f_{i0}^\alpha{\rm Im \;}h_i^{\alpha\prime}+{\rm Im\;}h_i^{\alpha*}h_i^{\alpha\prime}\pkt
\end{equation}
This relation can be used to simplify the
kinetic term in the quantum part of the
$tt$ component of the energy-momentum tensor. 
Using the expansion in terms of reduced mode functions and Eq. \eqn{eq:wronres}
one finds
\begin{equation}
\sum_\alpha\frac{{f_i^{\alpha}}'{f_i^{\alpha*}}'}{2\Omega_{\alpha0}}
=\sum_\alpha\frac{1}{2\Omega_{\alpha0}}
\left[\Omega_{\alpha0}^2(f_{i0}^{\alpha2}
-2f_{i0}^\alpha{\rm Re \;}h_i^\alpha-|h_i^\alpha|^2)
+|h_i^{\alpha\prime}|^2\right]\pkt
\end{equation}
 This can be combined with
\begin{eqnarray}
\sum_\alpha\frac{f_i^{\alpha}f_i^{\alpha*}}{2\Omega_{\alpha0}}
=\sum_\alpha\frac{1}{2\Omega_{\alpha0}}
\left[f_{i0}^{\alpha2}+2f_{i0}^\alpha{\rm Re \;} h_i^\alpha
+|h_i^\alpha|^2\right]
\end{eqnarray}
to obtain
\be\label{eq:kinetictermsimplified}
\sum_\alpha\frac{1}{2\Omega_{\alpha0}}
\left[{f_i^{\alpha}}'{f_i^{\alpha*}}'
+\Omega_{\alpha0}^2 f_i^\alpha f_i^{\alpha*}\right]
=\sum_\alpha\frac{1}{2\Omega_{\alpha0}}\left[2\Omega_{\alpha0}^2
f_{i0}^{\alpha2}+|h_i^{\alpha\prime}|^2\right]\pkt
\ee
When integrated over momentum the first term in the bracket, 
including the prefactor $1/2\Omega_{\alpha0}$ 
leads to a quartic divergence.
The second term is dominated at large momenta by the square of the
imaginary part of ${h_i^{\alpha(1)}}'$(see Eq. \eqn{eq:imh1as}):
\be
\frac{1}{2\Omega_{\alpha0}}|{h_i^{\alpha \,}}'|^2 \simeq
\frac{1}{2\Omega_{\alpha0}} {\rm Im\,}
\left({h_i^{\alpha \,}}'\right)^2
\simeq \sum_\beta \frac{1}{8  \Omega_{\alpha0}\Omega_{\beta 0}^2}
\tilde\calv_{\beta\alpha}\tilde\calv_{\alpha\beta}
\kma \ee
up to terms of order $k^{-4}$.
This implies a logarithmic divergence. Therefore the subleading terms
are not important, and we may replace this asymptotic estimate by
\be \label{eq:kineticdivergence}
\frac{1}{2\Omega_{\alpha0}}|{h_i^{\alpha \,}}'|^2 
\simeq \sum_\beta \frac{1}{4  \Omega_{\alpha0}\Omega_{\beta 0}
(\Omega_{\alpha0}+\Omega_{\beta 0})}
\tilde\calv_{\beta\alpha}\tilde\calv_{\alpha\beta}
\pkt \ee
This has the advantage that it can be integrated in closed form
and is of the same form as other logarithmically divergent
contribution, see Eq. \eqn{eq:logdivergence}.


\section{Large-momentum behaviour of a Fourier \\
transform}
\label{app:largemomentumbehavior}
\setcounter{equation}{0}  
When analysing the reduced mode function
$h^{\alpha (1)}_i$, Eq. \eqn{eq:h1},  at large momentum we encounter
the expression
\be
i \int_0^\tau d\tau'\calv_{jk}(\tau') 
\left[e^{i(\Omega_{\alpha0}+\Omega_{\beta0})(\tau-\tau')}-
e^{i(\Omega_{\alpha0}-\Omega_{\beta0})(\tau-\tau')}\right]\pkt
\ee
For $\alpha=\beta$ this integral has been analysed for large
$k$ in Ref.~\cite{Baacke:1996se}. \linebreak
For $\alpha\neq\beta$ 
the analysis is somewhat more involved.
The first exponential \linebreak
$\exp\left[i(\Omega_{\alpha0}+\Omega_{\beta0})(\tau-\tau')\right]$
will oscillate strongly when $k \to \infty$. So this part
of the integral
will be dominated by the region around $\tau=\tau'$, i.e., the
upper end of the integration interval. The large-momentum behaviour
can be analysed by repeated integrations by parts, as for the
case $\alpha=\beta$, see below. The second exponential
$\exp\left[i(\Omega_{\alpha0}-\Omega_{\beta0})(\tau-\tau')\right]$
will approach unity as $k \to \infty$.
Indeed 
\be
\Omega_{\alpha0}-\Omega_{\beta0}
= \frac{m_{\alpha0}^2-m_{\beta0}^2}{\Omega_{\alpha0}+\Omega_{\beta0}}
\simeq \frac{m_{\alpha0}^2-m_{\beta0}^2}{k}
\ee
as $k\to \infty$, so the exponent goes to zero in this limit.

Considering the two parts separately we find for the first one, via
integrations by parts,
\bea\nonumber
&&i \int_0^\tau d\tau'\calv_{jk}(\tau') 
e^{i(\Omega_{\alpha0}+\Omega_{\beta0})(\tau-\tau')}
\\\nonumber
&&=-\frac{1}{\Omega_{\alpha0}+\Omega_{\beta0}}\calv_{jk}(\tau)
\\&&+\frac{i}{(\Omega_{\alpha0}+\Omega_{\beta0})^2}
\left[\calv'_{jk}(\tau)-\calv'_{jk}(0)e^{i(\Omega_{\alpha0}+\Omega_{\beta0})\tau}\right] + \dots
\eea
and for the second one, by expanding the exponential,
\bea\nonumber
&&-i \int_0^\tau d\tau'\calv_{jk}(\tau') 
e^{i(\Omega_{\alpha0}-\Omega_{\beta0})(\tau-\tau')}
\\\nonumber&&=-i\int_0^\tau d\tau'\calv_{jk}(\tau')
-\frac{m_{\alpha0}^2-m_{\beta0}^2}{\Omega_{\alpha0}+\Omega_{\beta0}}
\int_0^\tau d\tau' \calv_{ij}(\tau')(\tau-\tau')
\\
&&+i\frac{(m_{\alpha0}^2-m_{\beta0}^2)^2}{2
(\Omega_{\alpha0}+\Omega_{\beta0})^2}
\int_0^\tau d\tau' \calv_{ij}(\tau')(\tau-\tau')^2 + \dots
\eea
When contracted with  $f_{j0}^\beta f_{k0}^\alpha$ the second term
on the right hand side does not contribute as $\tilde \calv_{\alpha\beta}
=f_{j0}^\beta f_{k0}^\alpha \calv_{jk}$ is symmetric in $\alpha$ and $\beta$.


\section{Dimensionally regulated integrals}
\label{app:dimensionallyregulatedintegrals}
\setcounter{equation}{0}
For the dimensional regularisation we need some identities ($D=4-\epsilon$):
\begin{eqnarray}
&&\hspace{-1cm}\int\frac{d^{D-1}k}{(2\pi)^{D-1}}\frac{1}{2\Omega_{\alpha0}\Omega_{\beta0}(\Omega_{\alpha0}+\Omega_{\beta0})}\nonumber\kma\\
&=&\frac{1}{16\pi^2}\left[L_\epsilon-\ln\frac{m_{\alpha0}^2}{\mu^2}+1+\frac{m_{\beta0}^2}{m_{\alpha0}^2-m_{\beta0}^2}\ln\frac{m_{\beta0}^2}{m_{\alpha0}^2}\right]\kma\\
&&\hspace{-1cm}\int\frac{d^{D-1}k}{(2\pi)^{D-1}}\frac{1}{2\Omega_{\alpha0}}=-\frac{m_{\alpha0}^2}{16\pi^2}\left[L_\epsilon-\ln\frac{m_{\alpha0}^2}{\mu^2}+1\right]\kma\\
&&\hspace{-1cm}\int\frac{d^{D-1}k}{(2\pi)^{D-1}}\Omega_{\alpha0}=-\frac{m_{\alpha0}^4}{32\pi^2}\left[L_\epsilon-\ln\frac{m_{\alpha0}^2}{\mu^2}+\frac{3}{2}\right]
\end{eqnarray}
where $L_{\epsilon}=\frac{2}{\epsilon}-\gamma+\ln 4\pi$.

In the limit $m_{\beta0}^2\rightarrow m_{\alpha0}^2$ the finite terms in 
the first of these equations reduce to
\begin{equation}
\lim_{m_{\beta0}^2\rightarrow m_{\alpha0}^2}\left[-\ln\frac{m_{\alpha0}^2}{\mu^2}+1+\frac{m_{\beta0}^2}{m_{\alpha0}^2-m_{\beta0}^2}\ln\frac{m_{\beta0}^2}{m_{\alpha0}^2}\right]=-\ln\frac{m_{\alpha0}^2}{\mu^2} \pkt
\end{equation}


\section{Counterterms}
\setcounter{equation}{0}
\label{sec:counterterms}

Using the explicit form of $\tilde\calm_{ij}^2$ the counterterm
Lagrangian Eq.~\eqn{eq:countertermlagrangian} can be decomposed 
into the usual mass, coupling constant and other counterterms as
\bea
\call^{{\rm ct}}
&=&-\frac{1}{2}\left( \delta m_i^2+ \delta \xi_i R\right)
a^2\tilde\phi_i^2 -
\frac{a^{4-n}}{4}\delta \lambda_{ij}\tilde \phi_i^2\tilde \phi_j^2
\\\nonumber
&&\hspace{20mm}+a^n \left(\frac{1}{2}\delta \tilde Z R + 
\frac{1}{2}\delta \tilde \alpha R^2 
-\delta \tilde\Lambda\right)\pkt
\eea
Note that the first terms do not have a prefactor $a^n$, as it has
been incorporated into the rescaled fields $\tilde \phi_i$.
\pagebreak

When we compare this to the expansion of the right hand side
of Eq.~\eqn{eq:countertermlagrangian} we find
\bea
\delta m_i^2 &=&\frac{1}{16 \pi^2}L_\epsilon
\left[3 \lambda_{ii}m_{i}^2+\sum_{j\neq i}\lambda_{ij}m_j^2\right]
\kma\\
\delta \xi_i &=&\frac{1}{16 \pi^2}L_\epsilon
 \left[3 \lambda_{ii}(\xi_i-1/6)+\sum_{j\neq i}\lambda_{ij}
(\xi_j-1/6)\right] 
\kma\\
\delta \lambda_{ii}&=&\frac{1}{16 \pi^2}L_\epsilon
\left[9 \lambda_{ii}^2+\sum_{j\neq i}\lambda_{ij}^2\right]
\kma\\
\delta\lambda_{ij}&=&\frac{1}{16 \pi^2}L_\epsilon
\lambda_{ij}\left[3\lambda_{ii}+3\lambda_{jj}
+4\lambda_{ij}\right]
\kma\\
\delta \tilde Z&=&-\frac{1}{16 \pi^2}L_\epsilon
\sum_i (\xi_i-1/6)m_i^2
\kma\\
\delta \tilde \alpha &=&-\frac{1}{32 \pi^2}L_\epsilon
\sum_i (\xi_i-1/6)^2
\kma\\
\delta \tilde \Lambda&=&\frac{1}{64 \pi^2}L_\epsilon
\sum_i m_i^4
\pkt\eea
These are in agreement with the counterterms found for the
case of a single field \cite{Baacke:1999gc}.
We recall that the last three renormalisation constants are 
related to renormalisation of the left-hand side
of Einstein's equations, which have been redefined as renormalisation
of the energy-momentum tensor, see Eqs.~\eqn{eq:Einsteinrenorm}
and \eqn{eq:energymomentumtensorredef}.

The counterterms do not by themselves define the renormalisation.
\linebreak
Choosing them as given above defines the $\overline{MS}$ scheme.
They can be mo\-di\-fied by finite terms in order that the
renormalised theory satisfies certain requirements.
E.g., if we want our universe to become matter or radiation dominated
at late times we will have to ensure that the effective
cosmo\-lo\-gical constant remains zero. We will also want
Newton's constant to retain its observed value.
The details  of such finite renormalisation
depend on the specific model and usually involve tedious
calculations. We do not address this issue here.


\section{An instructive example}
\setcounter{equation}{0}
\label{app:anexample}
The fluctuation integrals obviously play a central r\^ole in the 
one-loop appro\-xi\-mation to non-equilibrium dynamics.
In particular it plays a r\^ole for fixing the
parameters in the equilibrium theory, one would
not like the renormalisation conditions to be dependent
on the expansion of the universe. 
As we have mentioned already the $\ln a(\tau)$- term in the
finite parts $\calf_{ij}^{\rm ft}$ is new here and suggests a
logarithmic increase. It can become very large if the universe
expands by many $e$-foldings. So it is interesting to
analyse the behaviour of the subtracted integral $\calf_{ij}^{\rm sub}$
as well. We will do this for just one field, and setting
for simplicity $\xi=1/6$. Furthermore we will assume that the universe
has ended up, at least approximately, in its ground
state, i.e. that the classical field $\phi$ has reached its equilibrium value,
which we assume to be $0$ and that the fluctuations have 
almost died out, in a sense we will specify shortly. 

The mode equation for $f(k,\tau)$is given by
\be
f'' + (k^2 + \tilde \calm^2(\tau))f=0
\pkt\ee
If the classical field has reached its minimum $\calm^2$ is still
time dependent
\be
\tilde \calm^2(\tau)\simeq m^2 a^2
\kma\ee
so the modes are still under the influence of a time-depending
``potential''.
At least for large $k$ we can approximately solve the mode
equation using the semiclassical approximation
\be
f(k,\tau) \simeq {\rm C(k)}(k^2+\calm^2(\tau))^{-1/4}\exp( \pm i S)\kma
\ee 
with 
\be
S = \int_{\tau_0}^\tau d\tau' \sqrt{k^2+\tilde \calm^2(\tau')}
\pkt\ee
So we have
\be
|f(k)|^2 \simeq \frac{C^2(k)}{\sqrt{k^2+ \tilde \calm^2(\tau)}}
\pkt\ee
We will fix the constant $C(k)$ by the assumption that the system is,
at least as far as the high frequency modes are concerned,
already in its adiabatic vacuum at $\tau_0$. Then for
sufficiently large $k$
\be
C^2(k)\simeq \sqrt{k^2+\tilde\calm^2(\tau_0)}
\pkt\ee
 Then with $\tilde \calm^2
=m^2 a^2$ and $\calm^2(\tau_0)=m_0$ we have
\be
\calv(\tau)=m^2a^2(\tau)-m0^2\pkt 
\ee
and the integrand of subtracted fluctuation integral becomes
\be\label{eq:eikonalapproximation}
\frac{k^2}{4\pi^2}
\left[\frac{1}{\sqrt{k^2+m^2 a^2}}-
\frac{1}{\sqrt{k^2+m_0^2}}+\frac{ m^2a^2-m_0^2}{2\sqrt{k^2+m_0^2}^3}
\right]\pkt\ee
Integrating from $k=0$ to infinity we obtain
\be
\calf^{\rm sub}\simeq
\frac{1}{16\pi^2}\left(m^2a^2\ln \frac{a^2m^2}{m_0^2}  -m^2 a^2  + m_0^2\right)
\kma \ee
with corrections due to the low-energy modes.
The finite term $\calf^{\rm ft}$ here is given by
\be
\calf^{\rm ft}=\frac{1}{16\pi^2}\left
[m^2a^2\ln\frac{m_0^2}{\mu^2 a^2}- m_0^2
\right]\kma\ee
where we have used the single-field version of Eq. \eqn{eq:calfft}
For the finite value $\calf^{\rm fin}$ we therefore get
\be\label{eq:ffin}
\calf^{\rm fin}=\calf^{\rm ft}+\calf^{\rm sub}
\simeq\frac{m^2 a^2}{16\pi^2}\left(\ln\frac{m^2}{\mu^2}-1\right)
\pkt\ee
The logarithmic dependence on $a$ has disappeared. The factor 
$a^2$ is natural. In the equation of motion the fluctuation integral 
adds to $\tilde \phi^2= a^2 \phi^2$.

As an application of interest we may consider the broken symmetry case. 
We have a Higgs field $\chi$, with a classical
vacuum expectation value $v$. If its self-coupling is denoted
by $\alpha$ and if we use the $\overline{MS}$ scheme
the one-loop corrected equilibrium position is determined
by 
\be
< \tilde \chi^2> = v^2 a^2 + \alpha \calf^{\rm fin}
\pkt\ee
With Eq. \eqn{eq:ffin} we see that the expectation value of $\chi$ 
itself becomes independent of $a$, as it should. 
Were it not for the $\ln a$ term in $\calf^{\rm ft}$,
the vacuum expectation value would shift continuously with the expansion
of the universe.

Of course there are corrections, but this is to be expected.
In particular, if the fluctuations have thermalised, 
the vacuum expectation value of the Higgs field becomes a function 
of temperature. So our estimate would be strictly valid only after 
the temperature has reached $0$. 

We would like to add that the heuristic estimates for the
behaviour of the quantum fluctuations which we have used here 
are corroborated by numerical simulations. Actually they may
be used for estimating the high momentum
contribution to the subtracted fluctuation integral.

\bibliography{coupledfields}
\bibliographystyle{h-physrev4}
\end{document}